	\documentclass{aa}  
	\usepackage[colorlinks=true,citecolor=blue]{hyperref}
	\usepackage{xcolor}
	\hypersetup{
		colorlinks,
		linkcolor={red!50!black},
		citecolor={blue!50!black},
		urlcolor={blue!80!black}
	}
	\usepackage{caption}
	\usepackage{subcaption}
	\usepackage{booktabs}
	\usepackage[flushleft]{threeparttable}[hb]
	\usepackage[toc]{appendix} 
	\usepackage{graphicx}
	\usepackage{adjustbox}
	\usepackage{placeins}
	\usepackage{makecell}
	\usepackage{txfonts}
\begin{document}

	   \title{Co-moving groups around massive stars in the Nuclear Stellar Disk }
	
	
	   \author{Á. Martínez-Arranz
	   	\inst{1}
	   	\and
	   	R. Sch\"odel
	   	\inst{1}
	   	\and
	   	F. Nogueras-Lara
	   	\inst{2}
	   	\and M. W. Hosek Jr.
	   	\inst{3}
	   	\and F. Najarro
	   	\inst{4}
	   }
	   \institute{Instituto de Astrofísica de Andalucía (CSIC), University of Granada,
	   	Glorieta de la astronomía s/n, 18008 Granada, Spain\\
	   	\email{amartinez@iaa.es}
	   	\and
	   	European Southern Observatory, Karl-Schwarzschild-Strasse 2, 85748 Garching bei M\"unchen, Germany
	   	\and
	   	 University of California, Los Angeles, Department of Astronomy, Los Angeles, CA 90095, USA
	   	 \and
	   	 Centro de Astrobiología (CSIC/INTA), ctra. de Ajalvir km. 4, 28850 Torrejón de Ardoz, Madrid, Spain
	   }
	  \date{Received XXX; accepted YYY}

	 
	  \abstract
	   {During the last $\sim$ 30\,Myr the nuclear stellar disk in the Galactic center has been  the most prolific star forming region of the Milky Way when averaged by volume. Remarkably, the combined mass of the only three clusters present today in the nuclear stellar disk adds up to only $\sim$10\% of the total expected mass of young stars formed in this period. Several causes could explain this apparent absence of clusters and stellar associations. The stellar density in the area is so high that only the most massive clusters would be detectable against the dense background of stars. The extreme tidal forces reigning in the Galactic center could dissolve even the most massive of the clusters in just a few Myr. Close encounters with one of the massive molecular clouds, that are abundant in the nuclear stellar disk, can also rapidly  make any massive cluster or stellar association dissolve beyond recognition.  However, traces of some  dissolving young clusters/associations could still be detectable as co-moving groups.}
	   { It is our aim to identify so far unknown clusters or groups of young stars in the Galactic Center. We focus our search on known, spectroscopically identified massive young stars to see whether they can pinpoint such structures. }
	   {We created an algorithm to detect over-densities in the five-dimensional space spanned by proper-motion, position on the plane of the sky  and line-of-sight distances, using reddening as a proxy for the latter. Since co-moving groups must be young in this environment, proper motions provide a good means to search for young stars in the Galactic center. To this purpose we combined publicly available data from three different surveys of the Galactic center, covering an area of $\sim$\,160 arcmin$^2$ on the nuclear stellar disk.}
	   {We found four co-moving groups around massive stars, two of which are very close in position and velocity to the Arches'  most likely orbit. }
	   {These co-moving groups are strong candidates to be clusters or associations of recently formed stars, showing that not all the apparently isolated massive stars are run-away former members of any of the three known cluster in the Galactic center or simply isolated massive stars. Our simulations show that these groups or clusters may dissolve beyond our limits of detection in less than $\sim$6\,Myr.}
	
	
	   \maketitle
		\section{Introduction}
\begin{figure*}
			
			\includegraphics[width=\hsize]{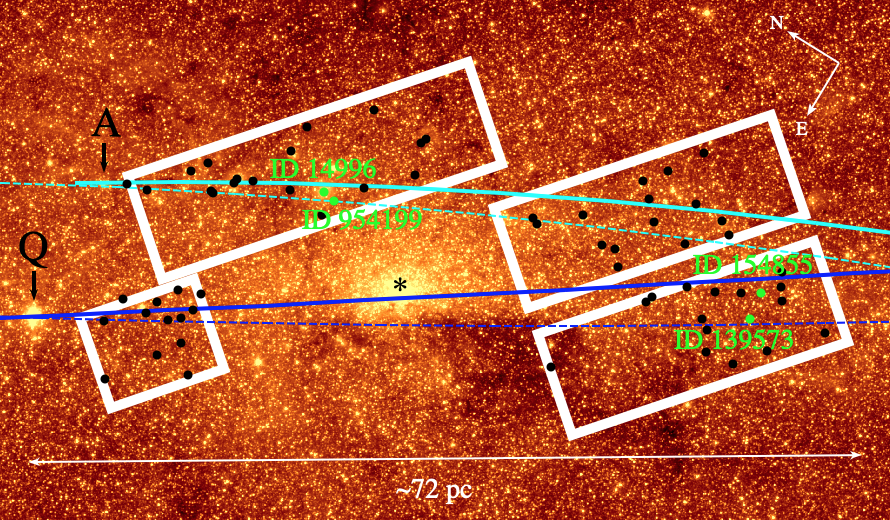}
			\caption{ Regions covered by \cite{LIBRALA2021} (indicated by white boxes) are superimposed on a 4.5 $\mu m$ Spitzer/IRAC image \citep{picture_2}. The cyan and blue lines represent the most probable prograde orbits of the Arches and Quintuplet clusters, respectively, as determined by \cite{Hosek2022}. The continuous line indicates movement towards the Galactic East (in front of the plane of the sky), while the dashed line indicates movement towards the Galactic West (behind the plane of the sky). The plotted points correspond to massive stars identified by \cite{Massive_stars} for which proper motion data is available. Among these stars, the green points are associated with a co-moving group, and their ID numbers correspond to their index as listed in \cite{LIBRALA2021}. The letters A and Q, along with the asterisk, denote the positions of the Arches and Quintuplet clusters and SgrA*.}
			
			\label{region}
		\end{figure*}

		Located around the Galactic center (GC), 8.2\,kpc away from Earth \citep{GC_distance_gravity}, we can find the Nuclear Stellar Disk (NSD), a flat-rotating structure \citep{Sch2015, Ban_catalog} of $\sim$200\,pc across and $\sim$50\,pc scale height \citep{NSD_size1, Laly_2020}. 
		The NSD is an old structure, with most of its stellar population at least as old as $\sim$\,8\,Gyr \citep{Paco_sfh}.\\	
			 The NSD constitutes an extreme environment marked by intense tidal forces, elevated stellar density, and exceptionally strong magnetic fields. Despite these challenging conditions, the NSD emits approximately 10\% of the total Lyman continuum flux in the entire Milky Way, while occupying less than 1\% of the galaxy's volume \citep{Lyman_flux_III,Lyman_flux_II,Lyman_flux}. Recent studies suggest that intense star forming activity occurred in the NSD between about 0.1 and 30 Myr ago,  reaching a star forming rate of about 0.1\,M$_{\odot}$ per year in this period \citep{three_cepheids, Paco_sfh}. This would correspond to more than one million solar masses of young stars. While such intensive star formation would leave clear signs in the form of massive stellar clusters and associations in the Milky Way's disk, the evidence for recent star formation in the NSD is more indirect. For example, there are only two known massive young clusters: the Arches and Quintuplet clusters, both at about 25\,pc projected distance from Sagittarius A*, and the association of young, massive stars in the central parsec \citep{Bartko_2010, Lu_2013}. They formed between 2-6 Myr ago and comprise about $1\times10^{4}\,$M$_{\odot}$ each. In addition, a few dozen massive young stars have been detected distributed throughout the central 100 pc \cite[e.g.][]{Massive_stars,Cano, Clark_2023}. Finally, on the order of $1\times10^{5}\,$M$_{\odot}$ of young stars of age $\sim$10 Myr have been reported to be present in the Sgr B1 HII region \citep{Paco_SgrB2}. Together, all these stars still make up only a fraction of the stars that formed in the past few tens of Myr. Where are the "missing" young stars? \\
			 This absence of direct observations of the products of star formation is due to the peculiar characteristics of the GC region. On the one hand, the stellar surface density is extremely high, which makes it hard to impossible to detect any but the most massive clusters in the form of local stellar over-densities. \\
			 On the other hand, extreme interstellar extinction and its variability on small angular scales means that young hot stars cannot be easily distinguished photometrically from cool, old giants \cite[see ][]{rainer_2014, Cano}. The extreme and differential extinction in the GC \citep[e.g.][]{extinction_los, paco_exctinction} limits observations to the near infrared wavelength range where it is impossible to identify young clusters in color-magnitude diagrams (CMDs), which are highly affected by the reddening \citep{GNSI}.\\
			  Also, strong tidal forces in the GC will dissolve a cluster as massive as the Arches in $\lesssim$\,10\,Myr \citep{dissolve_GC, cluster_dissolution} blending it with the background population. Spectroscopy needs to be performed at high angular resolution, which implies a very small field of view. Therefore, conducting spectroscopic searches is not a practical option due to the extensive time required to sample the entire region. However some clusters/stellar associations could still be detectable as co-moving groups, which is a detection method that has hardly been explored so far (with the exception of \citealt{Ban_cluster}).\\
		Several studies have shown how stellar kinematics can unveil different kinds of structures, such as open clusters in Gaia data \citep{Castro2018} or substructures in the Galactic plane of the Milky Way \citep{filaments_mw}. In the GC, stellar proper motions have been previously used to study the structure of the NSD (see for example \citealt{Ban_catalog,yo,Noguera_solo}). Membership probabilities and orbits for the Arches and Quintuplet clusters  have also been derived using proper motion analysis \citep{Hosek2022}.\\
			We have created a new method to reveal co-moving groups in the highly crowded environment of the GC. This tool is based on the \textit{DBSCAN} algorithm \citep{dbscan} and a similar version of it has been  previously used by \cite{Castro2018} to detect open clusters in Gaia DR2. In this case we looked for over-densities in a five-dimensional parameter space. In this paper we present four co-moving groups in the GC associated with four different massive stars (Fig.\ref{region}) identified by \cite{Massive_stars} .\\
			\section{Data}
		We used proper motion data from the catalog by \cite{LIBRALA2021} (from now on L21) acquired with WFC3/HST, combined with photometric data from the GALACTICNUCLEUS catalog \citep{GNSI, GNSII} acquired with HAWKI/VLT.  In order to test the cluster search algorithm we used proper motion catalogs for the Arches and the Quintuplet clusters by \cite{Hosek2022} and  extinction maps and catalogs in the H and Ks band by \cite{GNSIV}. 	
			\subsection{Proper motions}
		\label{sect_pm}	
\begin{figure}
			\hspace*{-0.3cm}      
			\includegraphics[width=\hsize]{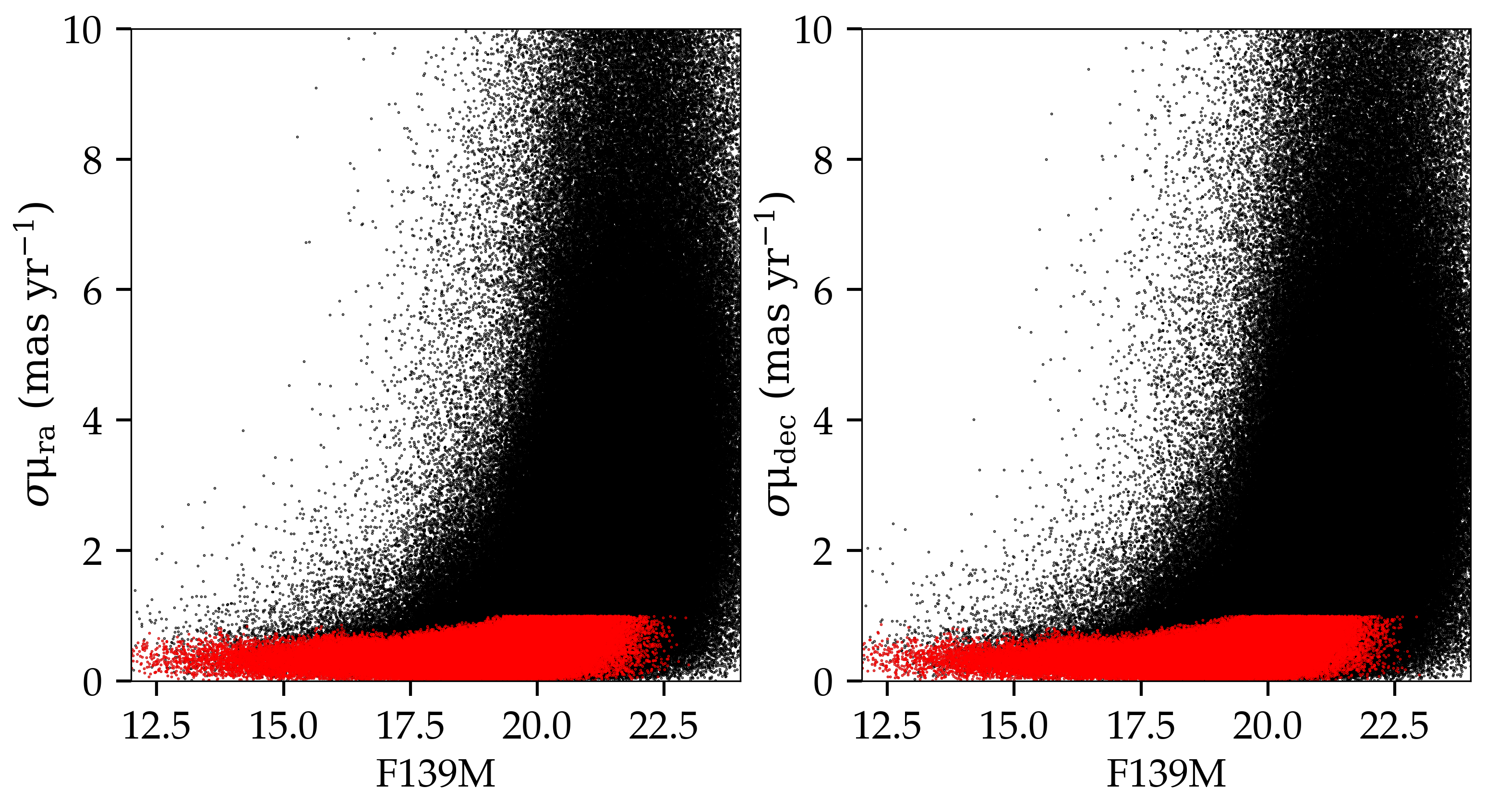}
			\caption{ Proper motion error in right ascension (left) and declination (right) versus F139M magnitude for the whole proper motion catalog (black) and the trimmed data (red).
			}
			\label{trimmdata}
		\end{figure}

		  The catalog of L21 was produced based on two sets of observations covering the area inside the white boxes in Fig.\,\ref{region}.  They were acquired with the NIR channel of the Wide-Field Camera 3, mounted on the HST, in October 2012 and August 2015. The proper motions were calibrated using reference stars for Gaia Data Release 2 \citep{Gaia2}. For more details about the acquisition, reduction and analysis of the data see L21.\footnote{The proper motions catalogs are available at \href{https://academic.oup.com/mnras/article/500/3/3213/5960177}{https://academic.oup.com/mnras/article/500/3/3213/5960177}}
			The final catalog consists of absolute proper motion measurement for $\sim$\,830.000 stars, that we trimmed in a similar way  as it was done by \cite{LIBRALA2021}, namely: we excluded stars with proper motions faster than 70\,mas\,/yr, we selected only stars with proper motion errors lower than the 85th percentile in bins of 0.1 mag width and, finally, we discarded stars with proper motion error bigger than 1\,mas\,/yr  (Fig.\,\ref{trimmdata}) .\\
			We cross-referenced the catalog with the GALACTICNUCLEUS survey \citep{GNSI,GNSII} to assign H and Ks magnitudes to the members of L21. GALACTICNUCLEUS was specifically designed to observe the GC, providing highly accurate point spread function photometry for over three million stars in the NSD and the innermost Galactic bar. The photometric uncertainties are remarkably low, remaining below 0.05 magnitudes at H$\sim$\,19 mag and Ks $\sim$\,18 mag. Once we obtained H and Ks magnitude values for the L21 members, we employed a color cut H$-$Ks > 1.3 to remove the foreground population.\\
			To assess the data quality, we extracted the mean velocity values for various components of the NSD and Bulge from L21 and compared them with values reported in the literature. Further information about this process can be found in Appendix \ref{quality_check}.
		
%

	\section{Methods}	
	\label{method}

	 We assumed that a stellar group would belong to the same cluster/stellar association if its members are close together in space and also have similar velocities. That is a six-dimensional parameter space, three dimensions for  the components of the velocities and three for the components of the position. We have available data only for proper motions and positions in the plane of the sky, but we can constrain indirectly the third dimension in position, i.e. the line-of-sight distance. Considering the considerable variation in extinction along the line of sight in the GC \citep[e.g.,][]{extinction_los, GNSI}, and the relatively constant intrinsic colors of the observable stard \citep[they vary by not more than a few 0.01 mag, see for example Fig.\,33 in][]{GNSI}, we hypothesize that changes in color are mainly influenced by extinction \citep[see also][]{GNSIV}. Therefore, if a group of stars shares similar colors, it is likely that they are located at a similar depth within the NSD \citep{Noguera_solo}. We search the data looking for over-densities in the five-dimensional space formed by proper motions along the Ra and Dec directions, coordinates in the plane of the sky and color.\\
	\subsection{The algorithm}
	We developed a tool for detecting co-moving groups in the GC based on DBSCAN (Density-Based Spatial Clustering of Applications with Noise) \citep{dbscan,dbscan_original,dbscan2}. DBSCAN  requires two input parameters: $\epsilon$ and $N_{min}$. The parameter $\epsilon$ establishes the distance within which the algorithm scans for nearby points around a specific data point. $N_{min}$ specifies the minimum number of points that should be within the $\epsilon$ radius to form a dense region. Considering these two parameters DBSCAN classifies each point in one of these 3 categories: \textit{Core point}, if the number of points around it within a radius of  $\epsilon$ is $\geq N_{min}$. \textit{Border Point}, if it is not a core point, but it is within an $\epsilon$ distance of one, and \textit{Noise}, if it is neither a core point nor a border point. The algorithm will iterate until all points are labeled with one of these categories. Core and border points are considered cluster members.\\
	The conditions present in the NSD, where the stellar densities vary greatly on scales of a few arcsec due to the high and patchy extinction \citep{GNSIV} and the high densities of stars \citep{GNSII}, make the selection of  $\epsilon$ particularly challenging. If we choose a value too small, the required minimum number of sources within a distance epsilon will never be fulfilled and no cluster will be found. On the other hand, if we choose a value too large, then spurious clusters present in the data just by chance, will be detected, because of statistical fluctuations.\\
	In order to find an appropriate value for $\epsilon$ we assumed that if there is a cluster in a particular data set then the distances among its members will be smaller, on average, than the distances between any other group of points in the same data set.  So, for each run of the algorithm we computed the distances to the $k^{th}\,nearest\,neighbor$ (k-NN) in the  five-dimensional space for all the stars in the area of analysis. Then, we generated a random sample with the same number of stars. To achieve this, we utilized the Gaussian kernel density estimator, specifically the \textit{gaussian\_kde} function from Scipy \citep{Scipy}, to estimate the distribution of each astrometric parameter from the original dataset. We sampled from the estimated distributions to create a simulated population. Subsequently, we computed the k-NN distances for the simulated population. Since these populations are randomly generated, any existing clusters present in the original data are effectively destroyed in the simulated population. To mitigate the inherent variability resulting from the random generation of simulations, we performed twenty different simulations and calculated the average values. This approach allowed us to minimize the impact of slight differences between individual simulations. \\
	If there were any cluster in the real data, then the minimum of the k-NN distances in the real data should be smaller than the minimum of the simulated data with  no cluster in it. Now, we choose the value for $\epsilon$ as the mean between both minima, the real and the simulated one\footnote{A similar method to constrain the value of $\epsilon$ were used by \cite{Castro2018}. }. By choosing an epsilon smaller than the minimum neighbor distance for the simulated data, we try to avoid any association of points that could show up in the data just by chance.\\
		\begin{figure}
		
		\includegraphics[width=\hsize]{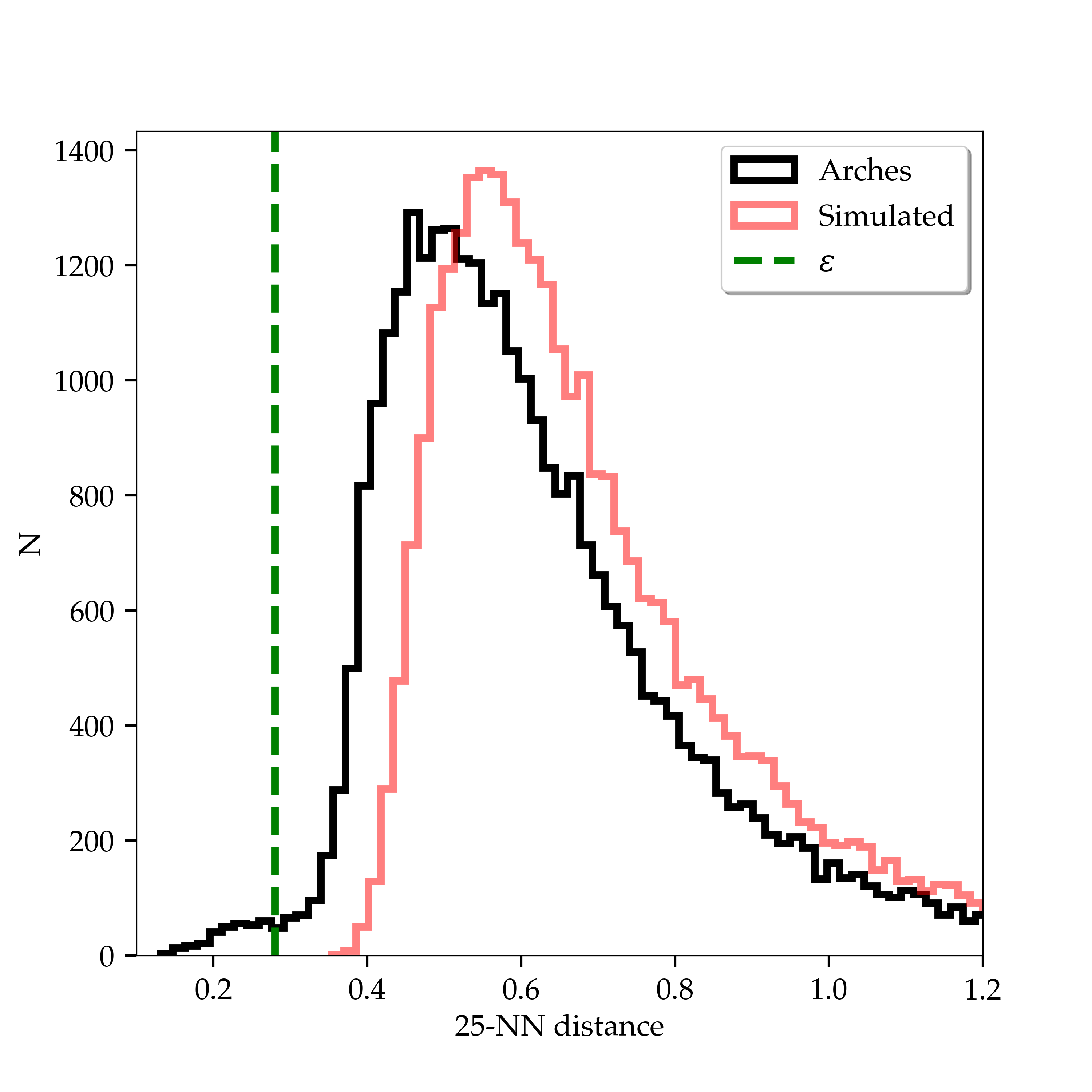}
		\caption{
			The black histogram represents the distance of each point, in the five dimensional space, to its 25th closest neighbor for the Arches data in H22. The red one is for the simulated data with no cluster in it. The green line marks the chosen value for epsilon  in this particular case.}
		\label{hist_arches}
	\end{figure}
	
	\subsection{Testing the algorithm}
	\label{testing}
\begin{figure*}
	  	\centering
	  	
	  	\begin{subfigure}[b]{1\textwidth}
	  		\hspace*{-0.3cm}      
	  		\includegraphics[width=\textwidth]{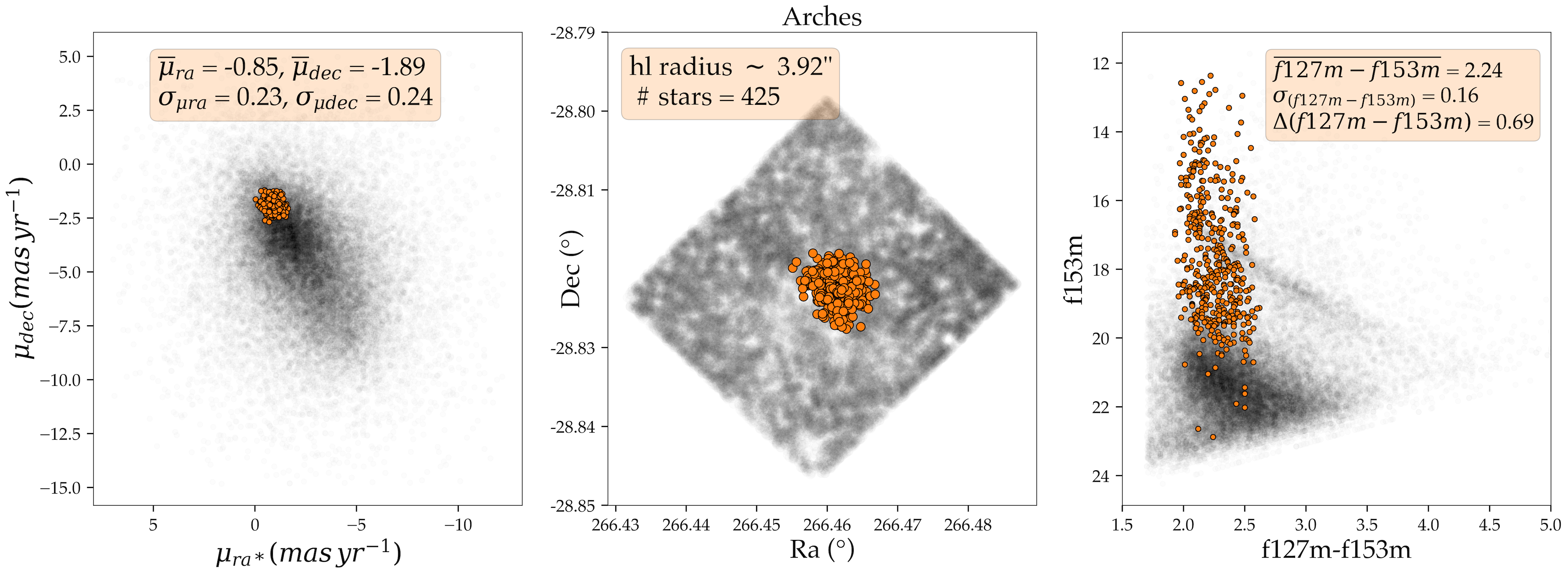}
	  	\end{subfigure}
	  	\hfill
	  	\begin{subfigure}[b]{1\textwidth}
	  		\hspace*{-0.3cm}      
	  		\includegraphics[width=\textwidth]{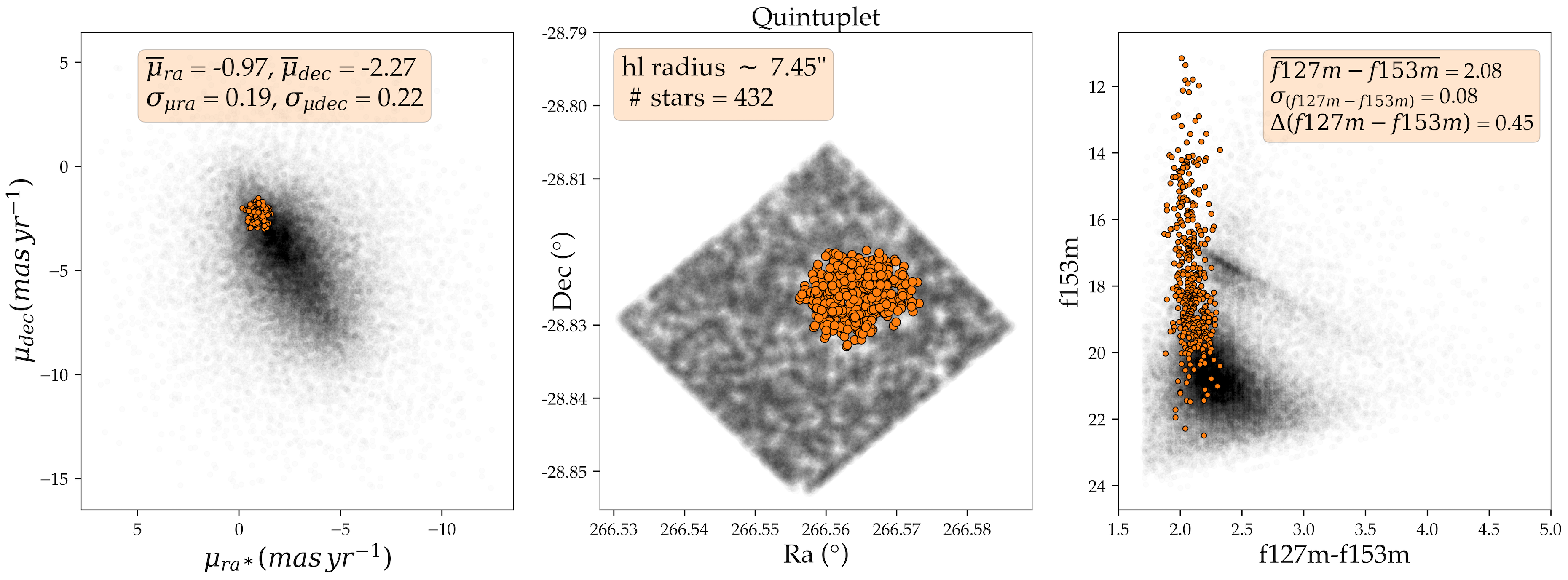}
	  	\end{subfigure}
	  	\caption{Arches (top row) and Quintuplet clusters as recovered by the algorithm in its 5D configuration. Left column: vector-point diagram. Middle column: Stellar positions. Right column: CMD. Orange points represent the objects labeled as cluster members. Black points represent objects that are not labeled as cluster members. }
	  	\label{hos_clusters}
	  \end{figure*}

\begin{figure}
		\includegraphics[width=\hsize]{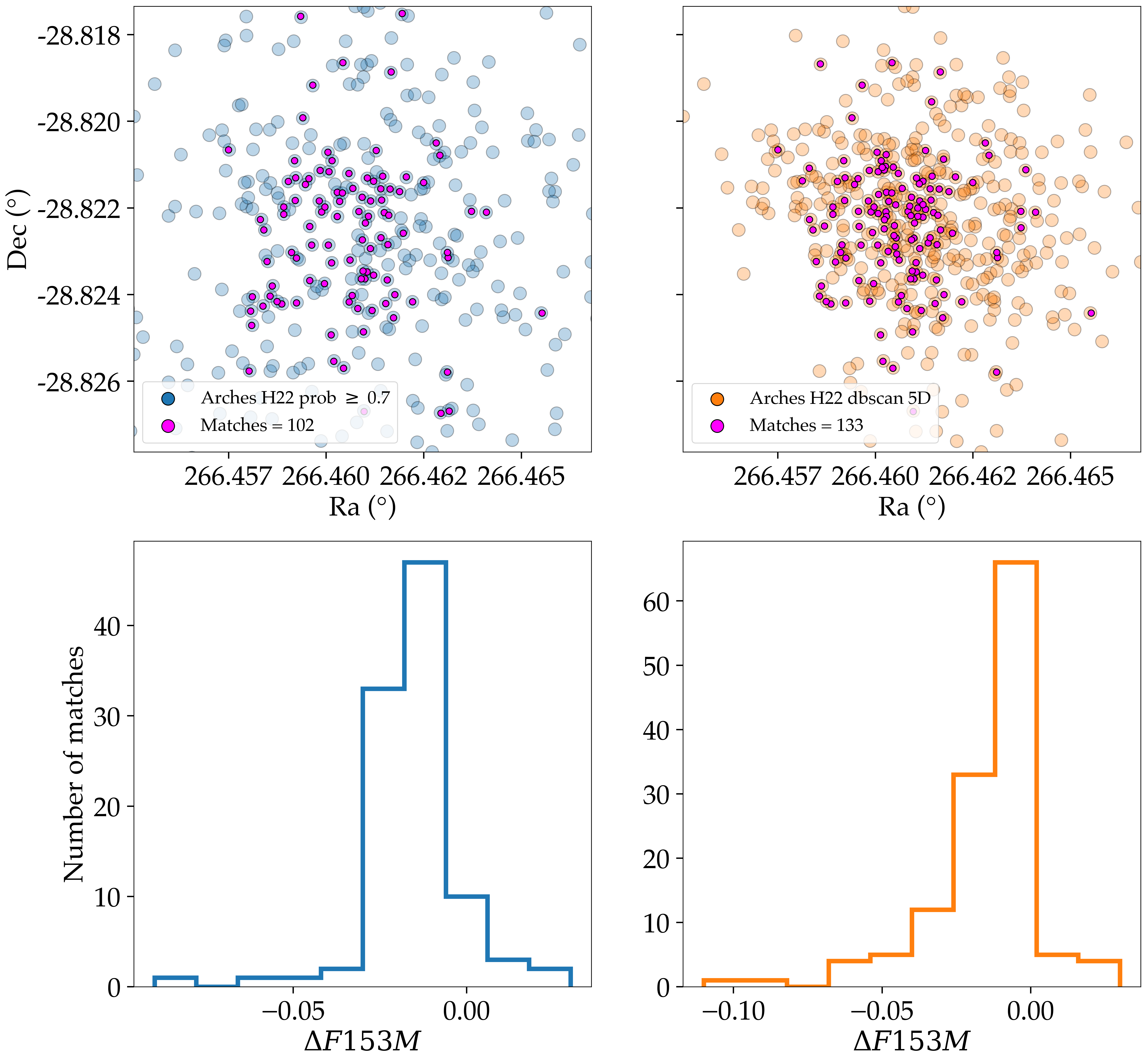}
		
		\caption{Top row: Blue points represent the Arches members according to H22, while orange points depict the ones selected by the algorithm in the 5D configuration. Fuchsia points represent the matches with the Arches members considered in C18 for each case. Bottom row: Magnitude residuals for the matches.}
		
		\label{clark_com}
	\end{figure}

	For testing our clustering tool we used data from \cite{Hosek2022} (from now on H22). They consist of astro-photometry data (equatorial coordinates, proper motions and magnitudes in F127M and F153M filters) acquired with the WFC3/HST camera in the areas of the Arches and Quintuplet clusters. In H22, membership probabilities for the Arches and Quintuplet clusters are assigned, considering stars as cluster members if their membership probability is greater than 0.7. Figure \ref{hos_prob} displays the stars identified as belonging to the Arches and Quintuplet clusters based on this criterion. For further insight into the probability assignment process, refer to Appendix B in \cite{Hosek2022}.\\
   In the following, we describe the processes we undertook, using the data for the Arches cluster in H22 as an example. First, we choose a starting value $N_{min}$ = 25. Then we computed the 25th-NN distance in a 5 dimensional space; velocity, position and color (black histogram in Fig. \ref{hist_arches}). Then we randomly generated a simulated population following the procedure described above, thus eliminating any real cluster from the data, and calculated the 25th-NN distance for the simulated data (red histogram in Fig. \ref{hist_arches}). We can see that the real set of data, that we know has a cluster in it, has smaller minimum 25-NN distances than the set of simulated data with no real cluster in it. These lower values correspond to the points that are closest in the 5D space.  Then we select our $\epsilon$ as the mean value between the minimum of the real data and the minimum of the simulated data (green dashed line in Fig. \ref{hist_arches}).\\
	 Now we run our algorithm on the Arches data set of H22. In Fig.\,\ref{hos_clusters} top row, we can see in orange the points labeled as cluster members that were returned by our algorithm. We repeated the process with the Quintuplet data set (bottom row in Fig.\,\ref{hos_clusters}). The mean values and their standard deviations for the proper motions that we obtain in each case are $(\mu_{*ra}, \mu_{dec})^{Arches}$= $-0.85$\,$\pm$\,0.23, $-1.89$\,$\pm$\,0.24\,mas/yr, and $(\mu_{*ra}, \mu_{dec})^{Quintuplet}$ = $-0.97$\,$\pm$\,0.19, $-2.29$\,$\pm$\,0.22\,mas/yr (left panels in Fig.\,\ref{hos_clusters}).
	 The values of the mean proper motions in both cases are similar to the ones obtained by \cite{Hosek2022}, i.e. $(\mu_{*ra}, \mu_{dec})^{Arches H22}$= $-0.80$, $-1.89$\,mas/yr, and $(\mu_{*ra}, \mu_{dec})^{Quintuplet H22}$ = $-0.96$, $-2.29$\,mas/yr. The velocity dispersions we obtained for both clusters, $\sigma \sim$\,0.2\,mas/yr, are comparable with the velocity dispersion found in other studies \citep{Arches_dispersion,Quintuplet_dispersion,Arches_dispersion_2}. We computed the half-light radii of both clusters by transforming the magnitudes into fluxes using the python package \textit{species} \citep{species}. These values are displayed in the orange boxes in the central plot of Fig.\ref{hos_clusters}. The ratio between these radii, approximately 2, aligns with the values reported in the literature for the half-light radius of the Arches cluster (12.5 arcsec) \citep{Arches_exctin} and the Quintuplet cluster (25 arcsec) \citep{Quintuplet_hl}. The smaller half-light radii values we obtained may indicate the detection limit of our algorithm, which appears to be less sensitive to the outer members of the clusters. 
	 When comparing the stars identified as members of the Arches and Quintuplet clusters by our algorithm with those that Hosek et al. (2022) considered as likely members of the clusters, we observe that approximately 55\% of the stars in the Arches cluster and around 65\% in the Quintuplet cluster coincide. These differences arise from restrictions in the parameter space of our algorithm. While the algorithm is configured to search for clusters in the 5D space, it also considers proximity in the parameter space defined by Ra and Dec coordinates as a requirement for a star to be considered a cluster member. Consequently, stars farther away from the cluster core, which are likely members according to \cite{Hosek2022} (Fig. \ref{hos_prob}, top row), are labeled as noise due to this criterion. If we relax the restrictions of the algorithm and search only in the parametric space of velocities, the coincident percentages increase to 74\% for the Arches and 85\% for the Quintuplet (Fig. \ref{hos_prob}, bottom row). However, it's important to note that due to the extreme crowding in the NSD and the fact that clusters as dense as the Arches or the Quintuplet are not expected to be found in the area, conducting a search for clusters or stellar associations using this configuration, which focuses solely on proximity in the velocity space, is not practical in the GC.\\
	We compare the Arches catalog calculated by \cite{Arches_stars} (hereafter C18) with the members identified in \cite{Hosek2022} (see \ref{hos_prob} top row, left plots) and those selected by our algorithm using the 5D configuration (top row of Fig.\,\ref{hos_clusters}). We display the matched positions in the top row of Fig. \ref{clark_com}. The C18 catalog comprises 194 stars, including confirmed and candidate Arches members. The matches between H22 and C18 are approximately 50\% of C18. In comparison, the percentage of matches with the algorithm-selected members is around 70\% of C18. This may indicate that the algorithm in its 5D configuration is effective at identifying members at the core of the clusters. In the bottom row of Fig.\,\ref{clark_com} , we present histograms of magnitude residuals for these matches. Given that the photometry in H22 and C18 originates from distinct catalogs, the low residual values with a mean of  $\overline{\Delta F153M}\sim$\,0.016, indicate non-spurious matches.\\
	 The Arches cluster experiences a significant variation in extinction, as discussed in the study by \cite{Arches_exctin}. This is evident in the broader distribution observed in the CMD of the stars identified as Arches members (Fig. \ref{hos_clusters}). We tried our algorithm with different values of $N_{min}$. We found that using any value of $N_{min}$ between 20 and 35  with H22 data, returned consistent results in radius, proper motions and velocities dispersion for both the Arches and Quintuplet datasets.

	We ran a second  test, consisting of randomly inserting the recovered Arches cluster by the algorithm in its 5D configuration (Fig. \ref{hos_clusters} top row), maintaining its original properties, into L21 (that does not contain Arches or Quintuplet) and then running the algorithm on the whole data set, L21 plus cluster.  We first crossmatched the data from H22 with the GALACTICNUCLEUS catalog \citep{GNSI,GNSII} in order to assign H and Ks magnitudes to the stars, in the same way as we did with L21. Since the extinction is not homogeneous across the NSD, we had to correct the color of the cluster stars according to the value of extinction at the place where the stars will be inserted. For this purpose we used the extinction maps in H and Ks from \cite{GNSIV}. We inserted and recovered the model cluster 50 times. The first and second rows in Tab.\,\ref{Table_disolved} show the mean motions and their dispersions for the inserted and for the recovered clusters. The last two columns of the table show the percentage of recovered stars and the percentage of contaminating stars that the recovered cluster contained. We recovered on average more than 80\% of the original stars with less than 20\% of contamination from other stars. The difference in $\mu_{*ra}$ and $\mu_{dec}$ between inserted and recovered cluster is $\sim$ 2\% 	\\
	Finally, in order to test the detection limit of the cluster algorithm, we repeated the experiment but this time we inserted less dense clusters. In order to simulate them we use as models the recovered cluster parameters for Arches and Quintuplet from the H22 data (Fig. \ref{hos_clusters}). Since the masses of both clusters are comparable, $\sim$ 10$^{4}$M$_{\odot}$ \citep{Arches_pm,Arches_mass} and they are in a similar environment, we assume that they will evolve in a similar way. We used Quintuplet, with an age of  $\sim$4\,Myr \citep{ Arches_Quintuplet_Ages, Quintuplet_4Myr,Quituplet_age} as a model for the evolutionary path that the younger Arches, $\sim$2.5 Myr \citep{Arches_Quintuplet_Ages,Arches_2Myr, Arches_2.5Myr}, would follow. By doing so, we can approximate the growth rate of the Arches cluster, assuming that its half-light radius will be similar to that of the Quintuplet in about 1.5\,Myr. Next, we move the Arches stars along the direction of their individual proper motion vectors, assuming this growth rate as constant over time. Then, we evolved the Arches cluster at different time lengths.\\
	 We inserted each of these models into the L21 data as we did before with the non-evolved models, and then ran our algorithm to recover them. This process was repeated 50 times for each model. The statistics for some of these simulations are presented in Table \ref{Table_disolved}. Given that the environment changes with each insertion, the table displays the average and standard deviation for the 50 insertions. We defined the detection limit of our algorithm when the percentage of model stars recovered by the algorithm became lower than the percentage of contaminating stars in the recovered cluster (last two columns in Table \ref{Table_disolved}). We can see that this limit is reached when our Arches model evolved  $\sim$\,3.3\,Myr.  
	 Based on this analysis, the detection of a hypothetical cluster as massive as the Arches, which has evolved over 6 million years since its detection, would exceed our detection limits, this is, more than 50\% of the members of this cluster would be probably contamination.This detection limit is comparable with theoretical predictions of the time it would take for a massive cluster to dissolve in the GC \citep{dissolve_GC,cluster_dissolution}.\\
	\begin{table*}
		\caption{Cluster recovering simulations}
		\begin{center}
			\begin{tabular}{cccccccc}
				\toprule
				Cluster  & $\mu_{ra}$ (mas/yr) & $\mu_{dec}$ (mas/yr) & $\sigma_{\mu ra}$ (mas/yr) & $\sigma_{\mu dec}$ (mas/yr) & R(arcsec) & \% recov. & \% contam. \\
				\midrule
				Model 0.0 Myr & -0.83 & -1.78 & 0.22 & 0.25 & 22 & - & - \\
				Recovered & -0.85$\pm$0.02 & -1.82$\pm$0.02 & 0.31$\pm$0.03 & 0.34$\pm$0.04 & 39$\pm$5 & 84$\pm$12 & 17$\pm$5 \\
				Model 0.6 Myr & -0.79 & -1.71 & 0.22 & 0.27 & 22 & - & - \\
				Recovered & -0.82$\pm$0.02 & -1.76$\pm$0.03 & 0.31$\pm$0.04 & 0.35$\pm$0.04 & 38$\pm$4 & 82$\pm$11 & 18$\pm$6 \\
				Model 1.5 Myr & -0.81 & -1.69 & 0.22 & 0.28 & 43 & - & - \\
				Recovered & -0.83$\pm$0.02 & -1.77$\pm$0.03 & 0.32$\pm$0.04 & 0.34$\pm$0.04 & 43$\pm$3 & 82$\pm$4 & 20$\pm$6 \\
				Model 2.1 Myr & -0.70 & -1.73 & 0.24 & 0.27 & 69 & - & - \\
				Recovered & -0.76$\pm$0.07 & -1.82$\pm$0.07 & 0.34$\pm$0.08 & 0.35$\pm$0.08 & 52$\pm$4 & 74$\pm$12 & 25$\pm$10 \\
				Model 3.0 Myr & -0.74 & -1.69 & 0.23 & 0.28 & 122 & - & - \\
				Recovered & -0.92$\pm$0.27 & -2.01$\pm$0.41 & 0.48$\pm$0.19 & 0.54$\pm$0.27 & 73$\pm$23 & 64$\pm$6 & 44$\pm$16 \\
				Model 3.3 Myr & -0.68 & -1.62 & 0.24 & 0.31 & 152 & - & - \\
				Recovered & -1.10$\pm$0.45 & -2.30$\pm$0.69 & 0.64$\pm$0.28 & 0.77$\pm$0.43 & 88$\pm$25 & 56$\pm$10 & 57$\pm$22 \\
				Model 3.6 Myr & -0.72 & -1.51 & 0.23 & 0.37 & 191 & - & - \\
				Recovered & -1.27$\pm$0.46 & -2.48$\pm$0.74 & 0.76$\pm$0.29 & 0.95$\pm$0.45 & 103$\pm$29 & 50$\pm$11 & 67$\pm$22 \\
				Model 3.9 Myr & -0.79 & -1.57 & 0.22 & 0.34 & 228 & - & - \\
				Recovered & -1.36$\pm$0.48 & -2.59$\pm$0.79 & 0.80$\pm$0.32 & 0.99$\pm$0.49 & 104$\pm$36 & 39$\pm$13 & 70$\pm$21 \\
				Model 4.5 Myr & -0.83 & -1.51 & 0.22 & 0.37 & 322 & - & - \\
				Recovered & -1.49$\pm$0.42 & -2.74$\pm$0.73 & 0.89$\pm$0.25 & 1.08$\pm$0.42 & 102$\pm$29 & 27$\pm$9 & 82$\pm$14 \\
				Model 5.1 Myr & -0.93 & -1.61 & 0.24 & 0.32 & 416 & - & - \\
				Recovered & -1.74$\pm$0.36 & -3.16$\pm$0.64 & 1.03$\pm$0.18 & 1.24$\pm$0.37 & 121$\pm$28 & 19$\pm$4 & 93$\pm$5 \\
				\bottomrule
			\end{tabular}
	\begin{tablenotes}
		\small
		\item Rows that start with the word \textit{Model} refer to the clusters we inserted to be recovered for the algorithm and the time we evolved then. The rest of rows refers to the statistics of the recovered clusters. Columns are: proper motion in Ra and Dec directions, standard deviations for the proper motions, cluster radius, percentage of stars recovered from the original model and percentage of stars contamination in the recovered cluster. The uncertainties are the standard deviation for the 50 runs in each case.
	\end{tablenotes}

	\end{center}
	\label{Table_disolved}
	
	\end{table*}
	\subsection{Mass Estimation}
   \label{mass_estimation}	
	
\begin{figure}
		\includegraphics[width=\hsize]{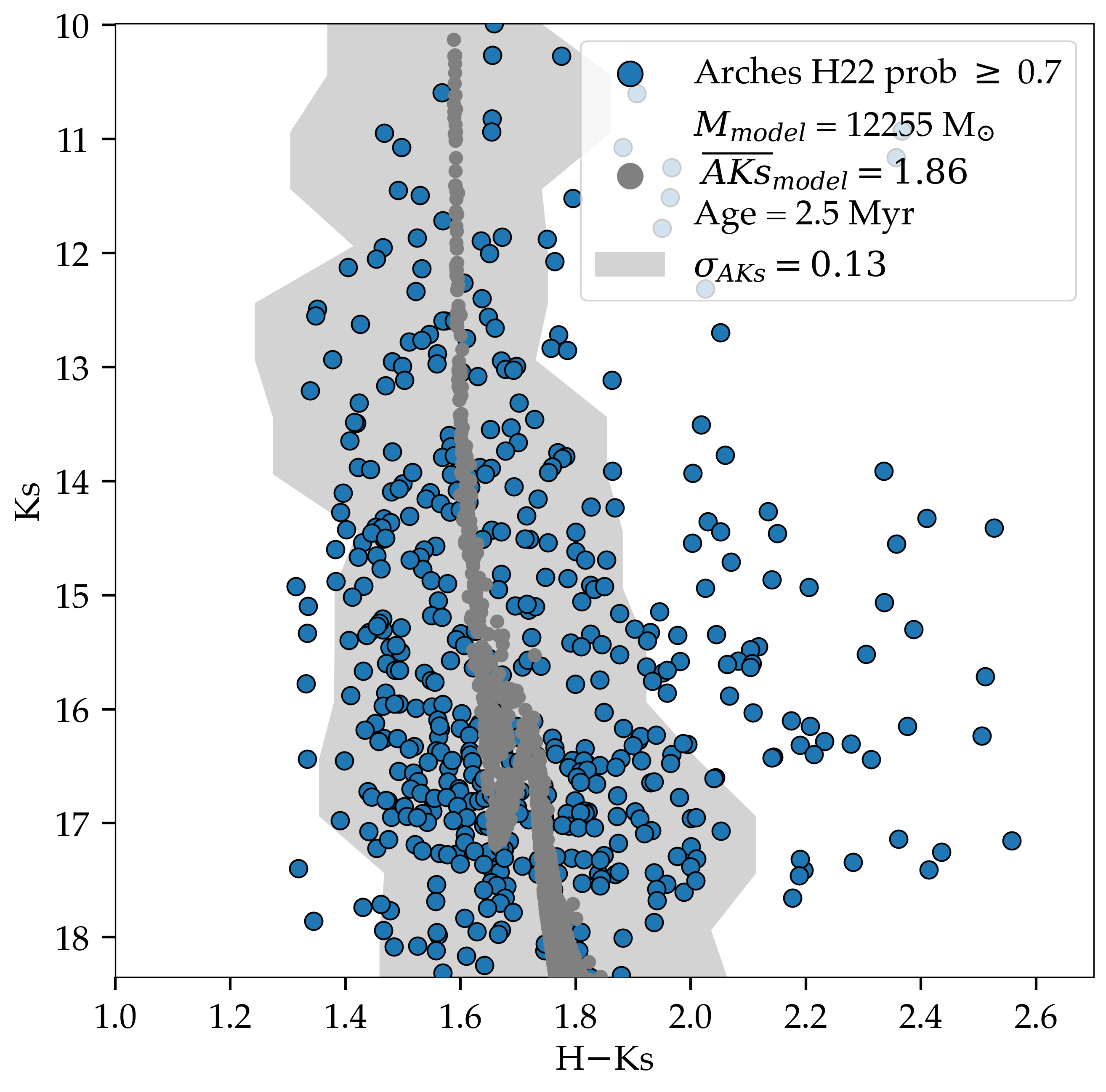}
		
		\caption{ The blue dots represent the members of the Arches cluster, while the gray dots represent a simulated cluster with an age of 2.5 Myr	. The mean extinction is given by AKs = 1.85. The shaded area in the plot represents the uncertainty in the position of the model members in the CMD due to differential extinction, $\sigma_{Aks}$.}
		
		\label{Arches_mass}
	\end{figure}

	In order to estimate the mass of the co-moving groups, we employed the python package \textit{Spisea} \citep{Spisea}. This package allows the generation of single-age, single-metallicity clusters, which we utilized to generate models for comparison with the selected co-moving groups. Firstly, we assigned extinction and differential extinction to the model. To compute these values, we utilized the extinction value of each star in the cluster from the catalog provided by \cite{GNSIV} along with its standard deviation.  Secondly, we assigned a mass and an age to the model and generated a simulated cluster. Then, we established a reference interval using a bright and a faint star within our co-moving group. Next, we compared the number of stars within this interval in our simulated cluster to that of the co-moving group. If the simulated cluster had a higher number of stars within the interval, we adjusted the mass of the model to a smaller value and generated a new simulated cluster. We repeated this process, gradually decreasing the mass of our simulation by 1\% increments until the number of stars in the reference interval of the model is not bigger than the number of stars inside the reference interval of the co-moving group.\\
	To assess the reliability of this approach, we initially applied this procedure using the members considered likely to be part of the Arches cluster according to \cite{Hosek2022}. (Fig. \ref{hos_prob} top row, left plots). The mean extinction was calculated by performing a crossmatch with the catalog for the GC by \cite{GNSIV}. In addition, we assigned an age of 2.5 million years to the model \citep{Arches_2Myr, Arches_2.5Myr} and solar metallicity  \citep{Arches_2Myr}. We adopted a one-segment power-law model for the Initial Mass Function (IMF) with a slope of $\alpha=-1.8$, according to \cite{Arches_IMF}. Due to the quality of proper motion data in the catalog, for magnitudes fainter than K$\sim$17 and brighter than K$\sim$11, the number of stars generated in the simulated cluster differs significantly (by more than threefold) from the observed count in the Arches cluster. As a result, we limited the reference interval for comparison to stars with magnitudes between K = 17 and K = 11 mag. Then, we iteratively adjusted the model mass until the number of stars in the model was not greater than the number of stars in the cluster. The procedure was repeated 50 times, and the resulting mean mass and standard deviation values were recorded. We obtained an estimated mass of approximately 12264 $\pm$ 495 M$_{\odot}$, which is consistent with the estimated mass of the Arches cluster  \citep{Arches_pm,Arches_mass}. The results of one of these 50 runs are presented in Fig. \ref{Arches_mass}.\\ 
	
			\newcommand{\IDstar}{14996}
				\newcommand{\IDmal}{427662}
			\section{Results}
\begin{figure}
 	\includegraphics[width=\hsize]{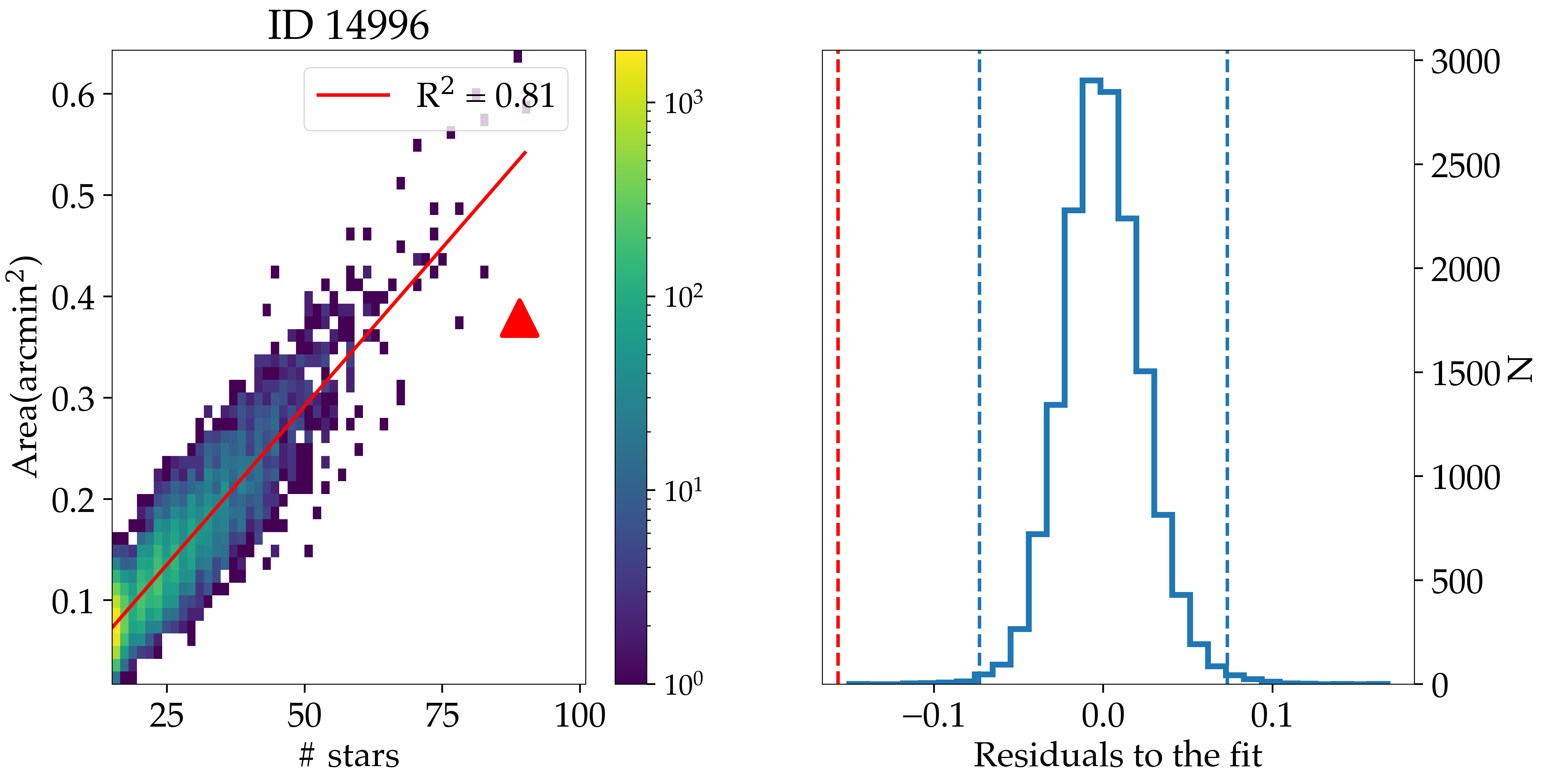}
 	
 	\caption{Left: Area versus the number of stars for $\sim$15000 statistical clusters identified by the algorithm across 10000 simulated populations. The red line represent the linear fit between the Area and the number of stars.The red triangle represents the co-moving group associated with star ID \IDstar. Right: The histogram represent the distribution of the residuals for the groups found in the simulated populations to the linear fit. The dashed blue lines mark the $\pm$3$\sigma$ levels of the distribution. The red dashed line mark the residual to the fit for the co-moving group associated with the stars ID \IDstar.
 	}

 	\label{ms_14996_simu}
 \end{figure}

\begin{figure}
 	\includegraphics[width=\hsize]{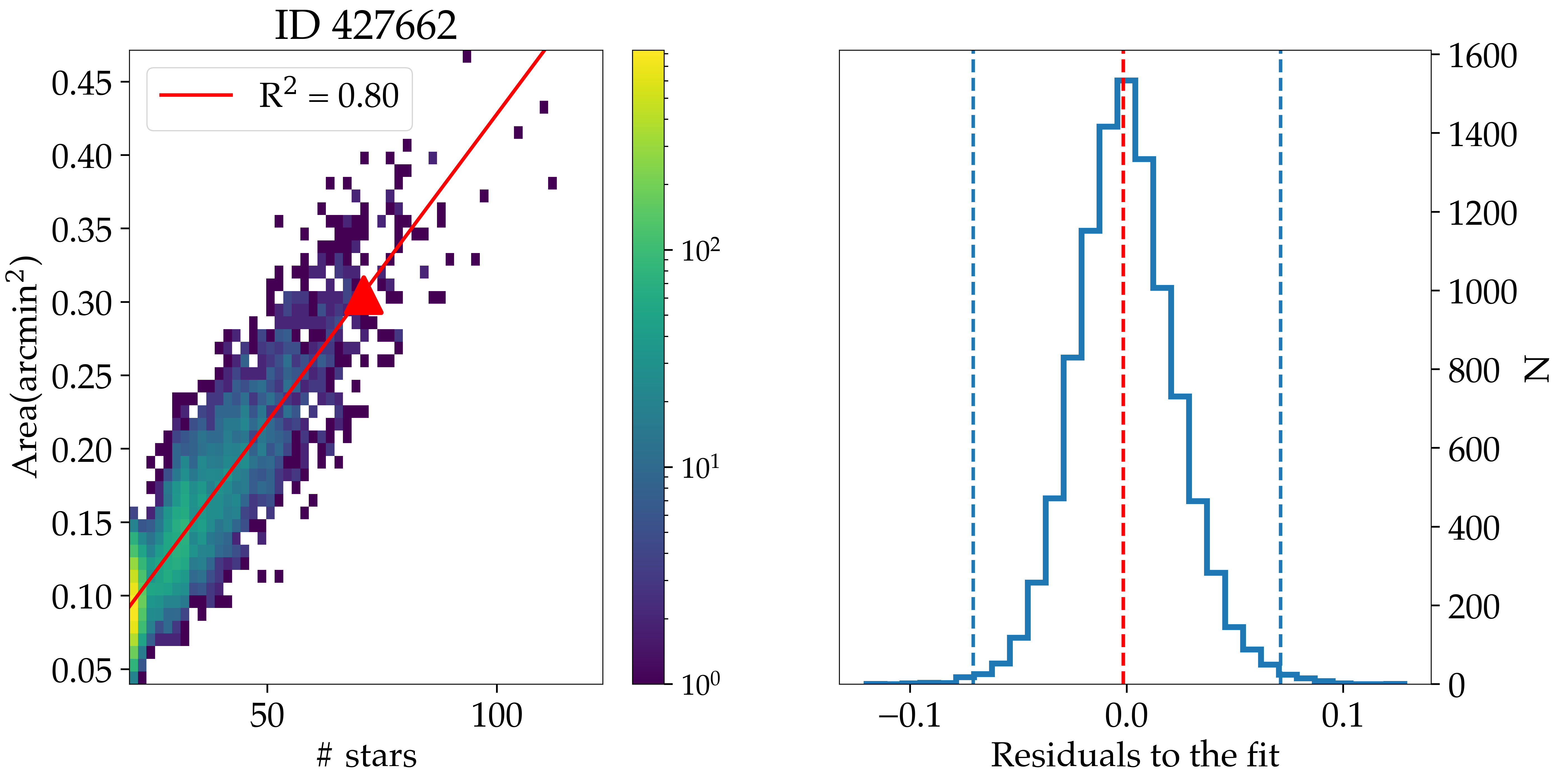}
 	
 	\caption{ Same as Fig.\ref{ms_14996_simu} for the  co-moving group associated with the star ID \IDmal
 	}
 	
 	\label{ms1059723_simu}
 \end{figure}

	Given the constraints posed by our algorithm and the relatively rapid dissolution of clusters in the NSD, we are compelled to confine our search to specific regions where the presence of young stars is likely. Thus, we restrict our search areas to regions within a radius from 50 to 150 arcsec around the massive stars listed in the catalog by \cite{Massive_stars}, which cannot be significantly older than 10 Myr. Then, we identify the massive young stars from \cite{Massive_stars} that have a counterpart in L21 after the quality cut. This selection accounts for a total of 59 objects, indicated by the black and green dots in Fig.\,\ref{region}. \\
	We divided the search methodology into two distinct parts. Firstly, we applied our algorithm to each of the massive stars, exploring 20 different configurations. These configurations involved changes in the search radius (50, 75, 100, 125 and 150 arcseconds) and the value for $N_{min}$ (15, 20, 25, 30). In this initial phase, we identified and selected six distinct massive stars that exhibit co-movement within a group.\\
The high density of sources in the NSD increases the probability of stars being closely positioned in the 5D space, which could potentially lead to the detection of spurious clusters or associations. 
To address this issue, we conducted a simulation-based study as the second part of our analysis.
 In this phase, we run our algorithm over simulated populations  and compared the resultant clusters with those obtained from real data. We conducted this analysis for each of the previously identified co-moving groups, utilizing the stars in their vicinity as the basis for the simulated populations. In the simulations, we kept the positions and magnitudes of the sources unchanged while randomly mixing their velocities. We then applied the algorithm to the simulated population. Since the velocities of the stars were shuffled, any group found in the simulations would represent statistical clusters, i.e., the outcome of random associations. We repeated this process 10000 times for each of the six cases. Subsequently, we compared the relationship between the area and the number of stars for these clusters with those found in the real data. The assigned area for each identified group corresponds to the minimum bounding box, calculated using the Python package \textit{alphashape}. We show as an example the results of this analysis for the co-moving groups associated with the stars ID \IDstar\, and \IDmal\, (Fig. \ref{ms_14996_simu} and Fig. \ref{ms1059723_simu}).  We can see in the left plots of these figures  that the area of the simulated clusters and the number of stars they contain exhibit a clear linear correlation. To quantitatively assess the likelihood that the co-moving group identified in the real data is merely a random association of stars, we compare it with the linear fit of the groups found in the simulations. Specifically, we compared the residual to the linear fit for the groups found in the simulations with the residuals to the same fit for the co-moving groups found in the real data. If the residuals of the co-moving group  identified in the real data does not surpass the 3$\sigma$ level of the residual distribution for the clusters found in the simulations, we discard it. Right plots in Fig. \ref{ms_14996_simu} and Fig. \ref{ms1059723_simu} illustrate this comparison for the co-moving group associated with the star ID \IDstar, which passed the cut, and ID \IDmal, which did not. We extended this analysis to all six identified co-moving groups linked to massive stars. Out of these, four groups successfully met the established criteria (green dots in Fig.\,\ref{region}) and were consequently considered unlikely to be the outcome of random stellar associations.
 
\begin{figure*}
  	\includegraphics[width=\hsize]{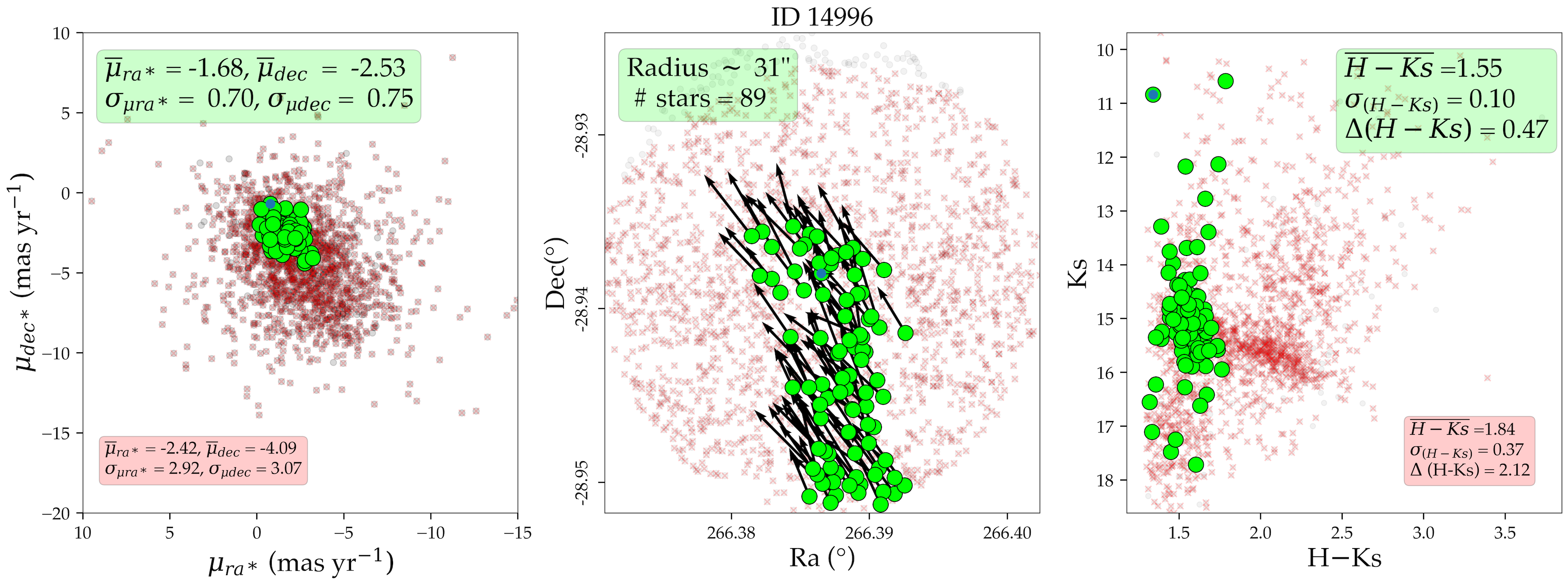}
  	
  	\caption{ 
  		From left to right, vector-point diagram, coordinates and CMD. Green points represent the members of the co-moving group, and the blue one the massive star. Arrows indicate the direction in the equatorial reference frame. Inside the green boxes are the values for the mean proper motion and sigma for the co-moving group associated with star ID \IDstar\, and number of stars members, the mean color and its sigma and the maximum difference in color within the group. The radius represents half the distance between the two farthest members of the group. Red crosses mark the stars in the neighborhood of the co-moving group and black dots the rests of the stars in the area. 	}
  	\label{ms_1}
  \end{figure*}

  Fig. \ref{ms_1} shows the vector point diagram, positions and CMD for the co-moving group associated with the star ID \IDstar.  Fig.\,\ref{ms_all} shows similar plots for the rest of the massive stars that are associated with a co-moving group that passed the final cut. In all four plots we show the co-moving groups with the smallest $\sigma_{\mu ra}$ and $\sigma_{\mu dec}$ that we found in each case. To provide a comparison, we have included red crosses to represent the stars surrounding the co-moving group within a distance of approximately 1.5 times the radius of the co-moving group. The corresponding values for these stars are displayed in the red boxes. The group associated with the massive star ID 14996 also contains the massive star ID 954199. Moreover, these two co-moving groups have 56 stars in common, which accounts for about 74\% of all members in the group associated with ID 954199. This high degree of overlap, combined with their similar velocities, suggests that these two groups could be part of a single, larger group.\\	
		\begin{figure}
		\includegraphics[width=\hsize]{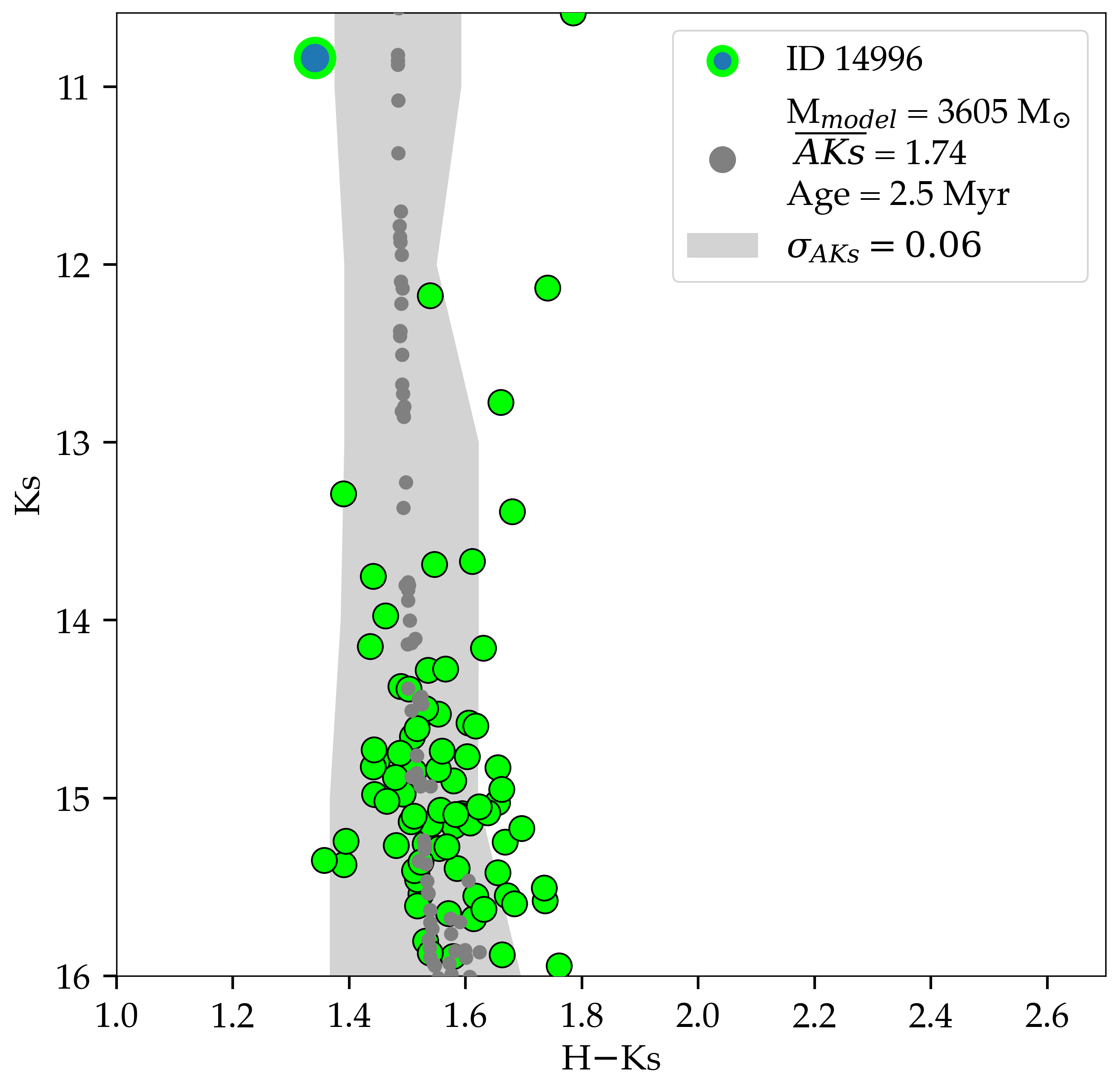}
		
		\caption{ Similar to Fig. \ref{Arches_mass} for the co-moving group associated with star ID 14996. The point with blue interior represents the massive star. }
		
		\label{14996_mass}
	\end{figure}	
    The four massive stars associated with a co-moving group are classified by \cite{Massive_stars} as either primary or secondary Paschen $\alpha$ emitters (Tab.\ref{Table_ms}), indicating that they are likely young stars. Furthermore, stars the with IDs 14996 and 154855 are reported by \cite{Ms_census} as a blue supergiant (O4-5 Ia$^{+}$) and a Wolf-Rayet (WN8-9ha) respectively. This classification suggests that they cannot be older than a few million years. \\
    The co-moving groups linked to stars IDs 14996 and 954199 are associated with a known HII region \citep{HII_regions} which further supports the presence of young stars. Both groups have a velocity comparable with the proper motions derived for the Arches cluster by \cite{LibralatoMC}, i.e. $\mu_{ra}* = -1.45\,\pm\,0.23$, $\mu_{dec}* = -2.68 \pm 0.14$ mas/yr, which are also calibrated with Gaia Data Release 2. Worth to mention that they lie along the path of  the most probable orbit for the Arches cluster calculated by \cite{Hosek2022} (cyan line in Fig. \ref{region}). The positions and velocities of these co-moving groups indicate that they may have formed in a similar location and possibly at a similar time as the Arches cluster. Another possibility is that these groups are part of the tidal tail resulting from the Arches cluster. The velocities and their projected distance from the Arches of approximately 20 pc align well with the tidal tail simulation presented in \cite{Arches_Quint_tails}.\\
   The mean extinction values and their standard deviations calculated for these clusters are $\overline{AKs}_{14996} = 1.74 \pm 0.06$ and $\overline{AKs}_{954199} = 1.76 \pm 0.05$. These values were derived using the extinction measurements for individual stars within each cluster from the catalog provided by \cite{GNSIV}. Notably, the Arches cluster, which experiences substantial differential extinction \citep{Arches_exctin}, demonstrates a standard deviation for the mean extinction of $\sigma_{AKs}$ = 0.13 (Fig.\,\ref{Arches_mass}). Conversely, the co-moving groups associated with ID \IDstar\, and 954199 display comparable and relatively low standard deviation values for their extinction. This suggests that these co-moving groups are not significantly affected by differential extinction, indicating that both clusters are located at a similar depth.\\
   It is worth mentioning that a co-moving group of six stars in the same area was identified in a different study by \cite{Ban_cluster}. This study used a different catalog and clustering method. Interestingly, three of these stars have a counterpart in the co-moving group associated with ID 14996. Additional investigations regarding this group will be presented in an upcoming publication (Martínez-Arranz et al. in prep.)\\
   Regarding the groups associated with stars ID 154855 and 139573, they display similar velocities and comparable mean extinctions: $\overline{AKs}_{154855} = 1.85 \pm 0.09$ and $\overline{AKs}_{139573} = 1.85 \pm 0.06$. Additionally, they are relatively close to each other in the plane of the sky. This proximity suggests that they may have been born as part of the same stellar formation process.\\
   The mean color for all four co-moving groups is around $\overline{H-Ks}$\,=\,1.55 (Fig.\,\ref{ms_1}, Fig.\,\ref{ms_all})). This indicate that they are located close to the outer edge of the NSD \citep[see Fig. 14 in] []{GNSII}.\\
	Given the numerous unknown parameters involved, such as cluster membership probability, age, metallicty, and IMF, estimating the masses of the co-moving groups presented in Fig. \ref{ms_1} and Fig. \ref{ms_all} becomes a challenging task. However, for the groups associated with massive stars ID \IDstar\, and 954199, if we consider the possibility that they are formed through the same process that gave rise to the Arches cluster or are part of its tidal tail, we can adopt similar assumptions for their IMF, metallicity, and age as used for the mass estimation of the Arches cluster (Fig.\,\ref{Arches_mass}). These assumptions include a top-heavy initial mass function \citep{Arches_IMF}, a solar metallicity \citep{Arches_2Myr} and an age of 2.5 Myr \citep{Arches_2.5Myr}. Following a procedure similar to the one described in section \ref{mass_estimation}, we have estimated the masses for the two groups, and the results are presented in the first two rows of Tab.\,\ref{Table_ms}. \\
	For the groups associated with the massive stars ID 154855 and 139573, we conducted a series of simulations using various combinations of metallicity ([M/H] = 0 and [M/H] = 0.3), age (2, 5, and 8 Myr), and two different IMF models: the broken power-law derived by \cite{IMF} and the top-heavy one derived by \cite{Arches_IMF}. This resulted in a total of 12 different combinations, each of which was run 50 times for both groups. The estimated masses, along with their standard deviations, are presented in Table\,\ref{Table_ms}.\\
	All four groups exhibit velocity dispersions ranging from 0.74 to 0.83 mas/yr. In our simulations (Tab.\,\ref{Table_disolved}), the clusters with these values of velocity dispersion show a contamination level around 55 to  65\%. If we assume a similar level of contamination in the co-moving groups we found, along with the aforementioned unknown parameters, it could potentially result in variations in the estimated masses by a factor of approximately 2.\\
	\begin{table}
		\caption{Massive stars within in a co-moving group.}
		\begin{center}
			\scalebox{0.95}{
		\begin{tabular}{ccc}
			\toprule
			ID and Type & Ra Dec & Mass (M$_{\odot}$) \\
			\midrule
			14996 Pp$\alpha$ & 17h45m32.7624s -28d56m16.67 & 3585 $\pm$ 425 \\
			954199  Sp$\alpha$ & 17h45m33.2952s -28d56m44.81 & 2682 $\pm$ 420 \\
			154855 Sp$\alpha$ & 17h45m09.6408s -29d11m30. & 5900 $\pm$ 850 \\
			139573 Pp$\alpha$ & 17h45m14.208s -29d11m41.50 & 5292 $\pm$ 825 \\
			\bottomrule
	\end{tabular}  }
			\begin{tablenotes}
				\small
				\item First column: ID for the massive stars as they appear in L21, and classification by \cite{Massive_stars}. Pp$\alpha$ and Sp$\alpha$ stand for primary and secondary Paschen $\alpha$ emitters. Second: coordinates. Third: estimated masses for the associated co-moving groups and their standard deviation.
			\end{tablenotes}
			
		\end{center}
		\label{Table_ms}
	\end{table}

			\section{Discussion and conclusions}
			
				We have developed a method to scan the GC  for co-moving groups that offers the possibility of tackling the so-called missing cluster problem under a new angle. We present here the first results of this new analysis, where we found four different co-moving groups around known massive stars in the NSD. Our toy model roughly estimates the time that it takes for a massive cluster in the GC to dissolve beyond the detection limit of our algorithm and, therefore, we are able to restrict the age of the co-moving groups that we present. We believe that the presence of these groups constitute a direct evidence of recent star formation in the GC. \\
				We analyzed the area around 59 known massive stars in the GC and found that four of them probably form part of a co-moving group. The relatively high velocity dispersion and low density of these co-moving groups, compared to those of the Arches or Quintuplet clusters, suggest two possible scenarios. Firstly, these co-moving groups may have originated from a dense cluster that has already undergone significant dissolution. Alternatively, they may have originated from a less dense stellar association. Recent studies have proposed that a substantial portion of the stars in the GC may have been born as part of loose associations of stars rather than gravitationally bound clusters \citep{cluster_formation_rate}. Supporting this scenario, the identification of $\sim$10$^{5}$\,M$_{\odot}$ of  young stars in the SgrB1 regions \citep{Paco_SgrB2}, which are only $\sim5$ million years older than the Arches and Quintuplet clusters, provides further evidence. In the specific case of the groups linked to the massive stars ID 14996 and 951499, there is a possibility that they are part of the tidal tail of the Arches cluster.\\
				  These groups show that not all apparently isolated massive stars in the NSD are run-away members from the Nuclear star cluster, the Arches or the Quintuplet, but highlight the location of stellar association\textbackslash clusters smaller than Arches or Quintuplet and or in an advanced state of dissolution \citep{Massive_stars}.\\
	            On the one hand, the small number of co-moving groups detected by our analysis may be influenced by the quality of the data set and by our conservative selection criteria. On the other hand, the large number of apparently unaccompanied massive young stars (along with the conclusion  by L21 that they are not runaways from the known massive clusters) provides evidence that massive stars may form in isolation in the GC. \\
				With the available data sets, we cannot estimate metallicities or radial velocities. Additionally, our estimations of ages and masses for the entire co-moving group are only rough approximations. To constrain these parameters and confirm the nature of these groups, future spectroscopy observations will be necessary.\\
			The proper motions catalog that we used in this paper covers only a fraction of the NSD and the uncertainty cut in proper motion that we made in the analysis significantly reduces the number of disposable sources. A wider and deeper set of data is necessary to continue with the search and corroborate these preliminary results. We are currently working on the reduction  of a second epoch of GALACTICNUCLEUS covering almost entirely the NSD. Combined with the first epoch \citep{GNSI} will result in an unprecedented  level of precision for proper motion measurements. Preliminary tests suggest an estimated uncertainty of $\sim$\,0.5\,mas/yr. \\
				This new technique opens exciting possibilities of research in the GC. A more complete detection of young clusters in the NSD would allow us to address the crucial  question of whether the IMF in the GC is fundamentally different from that in the Galactic disk.
			\begin{acknowledgements}	
				Author Á. Martínez-Arranz and 	R. Sch\"odel acknowledge financial support from the Severo Ochoa grant CEX2021-001131-S funded by MCIN/AEI/ 10.13039/501100011033. Á. Martínez-Arranz and R. Sch\"odel acknowledge  support from grant EUR2022-134031 funded by MCIN/AEI/10.13039/501100011033 and by the European Union NextGenerationEU/PRTR. and by grant  PID2022-136640NB-C21 funded by MCIN/AEI 10.13039/501100011033 and by the European Union.		We extend our gratitude to Paloma for generously sharing her expertise and guidance.
			\end{acknowledgements}
			
		\bibliographystyle{aa} 
	\bibliography{Lib_cit}

\begin{thebibliography}{65}
\expandafter\ifx\csname natexlab\endcsname\relax\def\natexlab#1{#1}\fi

\bibitem[{{Astropy Collaboration} {et~al.}(2022){Astropy Collaboration},
  {Price-Whelan}, {Lim}, {Earl}, {Starkman}, {Bradley}, {Shupe}, {Patil},
  {Corrales}, {Brasseur}, {N{\"o}the}, {Donath}, {Tollerud}, {Morris},
  {Ginsburg}, {Vaher}, {Weaver}, {Tocknell}, {Jamieson}, {van Kerkwijk},
  {Robitaille}, {Merry}, {Bachetti}, {G{\"u}nther}, {Aldcroft},
  {Alvarado-Montes}, {Archibald}, {B{\'o}di}, {Bapat}, {Barentsen},
  {Baz{\'a}n}, {Biswas}, {Boquien}, {Burke}, {Cara}, {Cara}, {Conroy},
  {Conseil}, {Craig}, {Cross}, {Cruz}, {D'Eugenio}, {Dencheva}, {Devillepoix},
  {Dietrich}, {Eigenbrot}, {Erben}, {Ferreira}, {Foreman-Mackey}, {Fox},
  {Freij}, {Garg}, {Geda}, {Glattly}, {Gondhalekar}, {Gordon}, {Grant},
  {Greenfield}, {Groener}, {Guest}, {Gurovich}, {Handberg}, {Hart},
  {Hatfield-Dodds}, {Homeier}, {Hosseinzadeh}, {Jenness}, {Jones}, {Joseph},
  {Kalmbach}, {Karamehmetoglu}, {Ka{\l}uszy{\'n}ski}, {Kelley}, {Kern},
  {Kerzendorf}, {Koch}, {Kulumani}, {Lee}, {Ly}, {Ma}, {MacBride}, {Maljaars},
  {Muna}, {Murphy}, {Norman}, {O'Steen}, {Oman}, {Pacifici}, {Pascual},
  {Pascual-Granado}, {Patil}, {Perren}, {Pickering}, {Rastogi}, {Roulston},
  {Ryan}, {Rykoff}, {Sabater}, {Sakurikar}, {Salgado}, {Sanghi}, {Saunders},
  {Savchenko}, {Schwardt}, {Seifert-Eckert}, {Shih}, {Jain}, {Shukla}, {Sick},
  {Simpson}, {Singanamalla}, {Singer}, {Singhal}, {Sinha}, {Sip{\H{o}}cz},
  {Spitler}, {Stansby}, {Streicher}, {{\v{S}}umak}, {Swinbank}, {Taranu},
  {Tewary}, {Tremblay}, {de Val-Borro}, {Van Kooten}, {Vasovi{\'c}}, {Verma},
  {de Miranda Cardoso}, {Williams}, {Wilson}, {Winkel}, {Wood-Vasey}, {Xue},
  {Yoachim}, {Zhang}, {Zonca}, \& {Astropy Project Contributors}}]{astropy}
{Astropy Collaboration}, {Price-Whelan}, A.~M., {Lim}, P.~L., {et~al.} 2022,
  \apj, 935, 167

\bibitem[{{Bartko} {et~al.}(2010{\natexlab{a}}){Bartko}, {Martins}, {Trippe},
  {Fritz}, {Genzel}, {Ott}, {Eisenhauer}, {Gillessen}, {Paumard}, {Alexander},
  {Dodds-Eden}, {Gerhard}, {Levin}, {Mascetti}, {Nayakshin}, {Perets},
  {Perrin}, {Pfuhl}, {Reid}, {Rouan}, {Zilka}, \& {Sternberg}}]{Bartko_2010}
{Bartko}, H., {Martins}, F., {Trippe}, S., {et~al.} 2010{\natexlab{a}}, \apj,
  708, 834

\bibitem[{{Bartko} {et~al.}(2010{\natexlab{b}}){Bartko}, {Martins}, {Trippe},
  {Fritz}, {Genzel}, {Ott}, {Eisenhauer}, {Gillessen}, {Paumard}, {Alexander},
  {Dodds-Eden}, {Gerhard}, {Levin}, {Mascetti}, {Nayakshin}, {Perets},
  {Perrin}, {Pfuhl}, {Reid}, {Rouan}, {Zilka}, \& {Sternberg}}]{IMF_topheavy}
{Bartko}, H., {Martins}, F., {Trippe}, S., {et~al.} 2010{\natexlab{b}}, \apj,
  708, 834

\bibitem[{{Cano-Gonz{\'a}lez} {et~al.}(2021){Cano-Gonz{\'a}lez}, {Sch{\"o}del},
  \& {Nogueras-Lara}}]{Cano}
{Cano-Gonz{\'a}lez}, M., {Sch{\"o}del}, R., \& {Nogueras-Lara}, F. 2021, \aap,
  653, A37

\bibitem[{{Castro-Ginard} {et~al.}(2018){Castro-Ginard}, {Jordi}, {Luri},
  {Julbe}, {Morvan}, {Balaguer-N{\'u}{\~n}ez}, \& {Cantat-Gaudin}}]{Castro2018}
{Castro-Ginard}, A., {Jordi}, C., {Luri}, X., {et~al.} 2018, \aap, 618, A59

\bibitem[{{Clark} {et~al.}(2018{\natexlab{a}}){Clark}, {Lohr}, {Najarro},
  {Dong}, \& {Martins}}]{Arches_stars}
{Clark}, J.~S., {Lohr}, M.~E., {Najarro}, F., {Dong}, H., \& {Martins}, F.
  2018{\natexlab{a}}, \aap, 617, A65

\bibitem[{{Clark} {et~al.}(2023){Clark}, {Lohr}, {Najarro}, {Patrick}, \&
  {Ritchie}}]{Clark_2023}
{Clark}, J.~S., {Lohr}, M.~E., {Najarro}, F., {Patrick}, L.~R., \& {Ritchie},
  B.~W. 2023, \mnras, 521, 4473

\bibitem[{{Clark} {et~al.}(2018{\natexlab{b}}){Clark}, {Lohr}, {Patrick},
  {Najarro}, {Dong}, \& {Figer}}]{Quituplet_age}
{Clark}, J.~S., {Lohr}, M.~E., {Patrick}, L.~R., {et~al.} 2018{\natexlab{b}},
  \aap, 618, A2

\bibitem[{{Clark} {et~al.}(2021){Clark}, {Patrick}, {Najarro}, {Evans}, \&
  {Lohr}}]{Ms_census}
{Clark}, J.~S., {Patrick}, L.~R., {Najarro}, F., {Evans}, C.~J., \& {Lohr}, M.
  2021, \aap, 649, A43

\bibitem[{{Clarkson} {et~al.}(2012{\natexlab{a}}){Clarkson}, {Ghez}, {Morris},
  {Lu}, {Stolte}, {McCrady}, {Do}, \& {Yelda}}]{Arches_dispersion_2}
{Clarkson}, W.~I., {Ghez}, A.~M., {Morris}, M.~R., {et~al.} 2012{\natexlab{a}},
  \apj, 751, 132

\bibitem[{{Clarkson} {et~al.}(2012{\natexlab{b}}){Clarkson}, {Ghez}, {Morris},
  {Lu}, {Stolte}, {McCrady}, {Do}, \& {Yelda}}]{Arches_pm}
{Clarkson}, W.~I., {Ghez}, A.~M., {Morris}, M.~R., {et~al.} 2012{\natexlab{b}},
  \apj, 751, 132

\bibitem[{{Dong} {et~al.}(2017){Dong}, {Lacy}, {Sch{\"o}del}, {Nogueras-Lara},
  {Gallego-Calvente}, {Mauerhan}, {Wang}, {Cotera}, \&
  {Gallego-Cano}}]{HII_regions}
{Dong}, H., {Lacy}, J.~H., {Sch{\"o}del}, R., {et~al.} 2017, \mnras, 470, 561

\bibitem[{Dong {et~al.}(2011)Dong, Wang, Cotera, Stolovy, Morris, Mauerhan,
  Mills, Schneider, Calzetti, \& Lang}]{Massive_stars}
Dong, H., Wang, Q.~D., Cotera, A., {et~al.} 2011, Monthly Notices of the Royal
  Astronomical Society, 417, 114

\bibitem[{{Espinoza} {et~al.}(2009){Espinoza}, {Selman}, \&
  {Melnick}}]{Arches_2.5Myr}
{Espinoza}, P., {Selman}, F.~J., \& {Melnick}, J. 2009, \aap, 501, 563

\bibitem[{Ester {et~al.}(1996)Ester, Kriegel, Sander, \& Xu}]{dbscan}
Ester, M., Kriegel, H.-P., Sander, J., \& Xu, X. 1996, in Proceedings of the
  Second International Conference on Knowledge Discovery and Data Mining,
  KDD'96 (AAAI Press), 226--231

\bibitem[{{Figer} {et~al.}(1999){Figer}, {Kim}, {Morris}, {Serabyn}, {Rich}, \&
  {McLean}}]{Arches_Quintuplet_Ages}
{Figer}, D.~F., {Kim}, S.~S., {Morris}, M., {et~al.} 1999, \apj, 525, 750

\bibitem[{{Gaia Collaboration} {et~al.}(2016){Gaia Collaboration}, {Prusti},
  {de Bruijne}, {Brown}, {Vallenari}, {Babusiaux}, {Bailer-Jones}, {Bastian},
  {Biermann}, {Evans}, {Eyer}, {Jansen}, {Jordi}, {Klioner}, {Lammers},
  {Lindegren}, {Luri}, {Mignard}, {Milligan}, {Panem}, {Poinsignon},
  {Pourbaix}, {Randich}, {Sarri}, {Sartoretti}, {Siddiqui}, {Soubiran},
  {Valette}, {van Leeuwen}, {Walton}, {Aerts}, {Arenou}, {Cropper}, {Drimmel},
  {H{\o}g}, {Katz}, {Lattanzi}, {O'Mullane}, {Grebel}, {Holland}, {Huc},
  {Passot}, {Bramante}, {Cacciari}, {Casta{\~n}eda}, {Chaoul}, {Cheek}, {De
  Angeli}, {Fabricius}, {Guerra}, {Hern{\'a}ndez}, {Jean-Antoine-Piccolo},
  {Masana}, {Messineo}, {Mowlavi}, {Nienartowicz}, {Ord{\'o}{\~n}ez-Blanco},
  {Panuzzo}, {Portell}, {Richards}, {Riello}, {Seabroke}, {Tanga},
  {Th{\'e}venin}, {Torra}, {Els}, {Gracia-Abril}, {Comoretto},
  {Garcia-Reinaldos}, {Lock}, {Mercier}, {Altmann}, {Andrae}, {Astraatmadja},
  {Bellas-Velidis}, {Benson}, {Berthier}, {Blomme}, {Busso}, {Carry},
  {Cellino}, {Clementini}, {Cowell}, {Creevey}, {Cuypers}, {Davidson}, {De
  Ridder}, {de Torres}, {Delchambre}, {Dell'Oro}, {Ducourant}, {Fr{\'e}mat},
  {Garc{\'\i}a-Torres}, {Gosset}, {Halbwachs}, {Hambly}, {Harrison}, {Hauser},
  {Hestroffer}, {Hodgkin}, {Huckle}, {Hutton}, {Jasniewicz}, {Jordan},
  {Kontizas}, {Korn}, {Lanzafame}, {Manteiga}, {Moitinho}, {Muinonen},
  {Osinde}, {Pancino}, {Pauwels}, {Petit}, {Recio-Blanco}, {Robin}, {Sarro},
  {Siopis}, {Smith}, {Smith}, {Sozzetti}, {Thuillot}, {van Reeven}, {Viala},
  {Abbas}, {Abreu Aramburu}, {Accart}, {Aguado}, {Allan}, {Allasia},
  {Altavilla}, {{\'A}lvarez}, {Alves}, {Anderson}, {Andrei}, {Anglada Varela},
  {Antiche}, {Antoja}, {Ant{\'o}n}, {Arcay}, {Atzei}, {Ayache}, {Bach},
  {Baker}, {Balaguer-N{\'u}{\~n}ez}, {Barache}, {Barata}, {Barbier}, {Barblan},
  {Baroni}, {Barrado y Navascu{\'e}s}, {Barros}, {Barstow}, {Becciani},
  {Bellazzini}, {Bellei}, {Bello Garc{\'\i}a}, {Belokurov}, {Bendjoya},
  {Berihuete}, {Bianchi}, {Bienaym{\'e}}, {Billebaud}, {Blagorodnova},
  {Blanco-Cuaresma}, {Boch}, {Bombrun}, {Borrachero}, {Bouquillon}, {Bourda},
  {Bouy}, {Bragaglia}, {Breddels}, {Brouillet}, {Br{\"u}semeister},
  {Bucciarelli}, {Budnik}, {Burgess}, {Burgon}, {Burlacu}, {Busonero}, {Buzzi},
  {Caffau}, {Cambras}, {Campbell}, {Cancelliere}, {Cantat-Gaudin}, {Carlucci},
  {Carrasco}, {Castellani}, {Charlot}, {Charnas}, {Charvet}, {Chassat},
  {Chiavassa}, {Clotet}, {Cocozza}, {Collins}, {Collins}, {Costigan}, {Crifo},
  {Cross}, {Crosta}, {Crowley}, {Dafonte}, {Damerdji}, {Dapergolas}, {David},
  {David}, {De Cat}, {de Felice}, {de Laverny}, {De Luise}, {De March}, {de
  Martino}, {de Souza}, {Debosscher}, {del Pozo}, {Delbo}, {Delgado},
  {Delgado}, {di Marco}, {Di Matteo}, {Diakite}, {Distefano}, {Dolding}, {Dos
  Anjos}, {Drazinos}, {Dur{\'a}n}, {Dzigan}, {Ecale}, {Edvardsson}, {Enke},
  {Erdmann}, {Escolar}, {Espina}, {Evans}, {Eynard Bontemps}, {Fabre},
  {Fabrizio}, {Faigler}, {Falc{\~a}o}, {Farr{\`a}s Casas}, {Faye}, {Federici},
  {Fedorets}, {Fern{\'a}ndez-Hern{\'a}ndez}, {Fernique}, {Fienga}, {Figueras},
  {Filippi}, {Findeisen}, {Fonti}, {Fouesneau}, {Fraile}, {Fraser}, {Fuchs},
  {Furnell}, {Gai}, {Galleti}, {Galluccio}, {Garabato}, {Garc{\'\i}a-Sedano},
  {Gar{\'e}}, {Garofalo}, {Garralda}, {Gavras}, {Gerssen}, {Geyer}, {Gilmore},
  {Girona}, {Giuffrida}, {Gomes}, {Gonz{\'a}lez-Marcos},
  {Gonz{\'a}lez-N{\'u}{\~n}ez}, {Gonz{\'a}lez-Vidal}, {Granvik}, {Guerrier},
  {Guillout}, {Guiraud}, {G{\'u}rpide}, {Guti{\'e}rrez-S{\'a}nchez}, {Guy},
  {Haigron}, {Hatzidimitriou}, {Haywood}, {Heiter}, {Helmi}, {Hobbs},
  {Hofmann}, {Holl}, {Holland}, {Hunt}, {Hypki}, {Icardi}, {Irwin}, {Jevardat
  de Fombelle}, {Jofr{\'e}}, {Jonker}, {Jorissen}, {Julbe}, {Karampelas},
  {Kochoska}, {Kohley}, {Kolenberg}, {Kontizas}, {Koposov}, {Kordopatis},
  {Koubsky}, {Kowalczyk}, {Krone-Martins}, {Kudryashova}, {Kull}, {Bachchan},
  {Lacoste-Seris}, {Lanza}, {Lavigne}, {Le Poncin-Lafitte}, {Lebreton},
  {Lebzelter}, {Leccia}, {Leclerc}, {Lecoeur-Taibi}, {Lemaitre}, {Lenhardt},
  {Leroux}, {Liao}, {Licata}, {Lindstr{\o}m}, {Lister}, {Livanou}, {Lobel},
  {L{\"o}ffler}, {L{\'o}pez}, {Lopez-Lozano}, {Lorenz}, {Loureiro},
  {MacDonald}, {Magalh{\~a}es Fernandes}, {Managau}, {Mann}, {Mantelet},
  {Marchal}, {Marchant}, {Marconi}, {Marie}, {Marinoni}, {Marrese},
  {Marschalk{\'o}}, {Marshall}, {Mart{\'\i}n-Fleitas}, {Martino}, {Mary},
  {Matijevi{\v{c}}}, {Mazeh}, {McMillan}, {Messina}, {Mestre}, {Michalik},
  {Millar}, {Miranda}, {Molina}, {Molinaro}, {Molinaro}, {Moln{\'a}r},
  {Moniez}, {Montegriffo}, {Monteiro}, {Mor}, {Mora}, {Morbidelli}, {Morel},
  {Morgenthaler}, {Morley}, {Morris}, {Mulone}, {Muraveva}, {Musella},
  {Narbonne}, {Nelemans}, {Nicastro}, {Noval}, {Ord{\'e}novic},
  {Ordieres-Mer{\'e}}, {Osborne}, {Pagani}, {Pagano}, {Pailler}, {Palacin},
  {Palaversa}, {Parsons}, {Paulsen}, {Pecoraro}, {Pedrosa}, {Pentik{\"a}inen},
  {Pereira}, {Pichon}, {Piersimoni}, {Pineau}, {Plachy}, {Plum}, {Poujoulet},
  {Pr{\v{s}}a}, {Pulone}, {Ragaini}, {Rago}, {Rambaux}, {Ramos-Lerate},
  {Ranalli}, {Rauw}, {Read}, {Regibo}, {Renk}, {Reyl{\'e}}, {Ribeiro},
  {Rimoldini}, {Ripepi}, {Riva}, {Rixon}, {Roelens}, {Romero-G{\'o}mez},
  {Rowell}, {Royer}, {Rudolph}, {Ruiz-Dern}, {Sadowski}, {Sagrist{\`a}
  Sell{\'e}s}, {Sahlmann}, {Salgado}, {Salguero}, {Sarasso}, {Savietto},
  {Schnorhk}, {Schultheis}, {Sciacca}, {Segol}, {Segovia}, {Segransan},
  {Serpell}, {Shih}, {Smareglia}, {Smart}, {Smith}, {Solano}, {Solitro},
  {Sordo}, {Soria Nieto}, {Souchay}, {Spagna}, {Spoto}, {Stampa}, {Steele},
  {Steidelm{\"u}ller}, {Stephenson}, {Stoev}, {Suess}, {S{\"u}veges}, {Surdej},
  {Szabados}, {Szegedi-Elek}, {Tapiador}, {Taris}, {Tauran}, {Taylor},
  {Teixeira}, {Terrett}, {Tingley}, {Trager}, {Turon}, {Ulla}, {Utrilla},
  {Valentini}, {van Elteren}, {Van Hemelryck}, {van Leeuwen}, {Varadi},
  {Vecchiato}, {Veljanoski}, {Via}, {Vicente}, {Vogt}, {Voss}, {Votruba},
  {Voutsinas}, {Walmsley}, {Weiler}, {Weingrill}, {Werner}, {Wevers},
  {Whitehead}, {Wyrzykowski}, {Yoldas}, {{\v{Z}}erjal}, {Zucker}, {Zurbach},
  {Zwitter}, {Alecu}, {Allen}, {Allende Prieto}, {Amorim},
  {Anglada-Escud{\'e}}, {Arsenijevic}, {Azaz}, {Balm}, {Beck}, {Bernstein},
  {Bigot}, {Bijaoui}, {Blasco}, {Bonfigli}, {Bono}, {Boudreault}, {Bressan},
  {Brown}, {Brunet}, {Bunclark}, {Buonanno}, {Butkevich}, {Carret}, {Carrion},
  {Chemin}, {Ch{\'e}reau}, {Corcione}, {Darmigny}, {de Boer}, {de Teodoro}, {de
  Zeeuw}, {Delle Luche}, {Domingues}, {Dubath}, {Fodor}, {Fr{\'e}zouls},
  {Fries}, {Fustes}, {Fyfe}, {Gallardo}, {Gallegos}, {Gardiol}, {Gebran},
  {Gomboc}, {G{\'o}mez}, {Grux}, {Gueguen}, {Heyrovsky}, {Hoar}, {Iannicola},
  {Isasi Parache}, {Janotto}, {Joliet}, {Jonckheere}, {Keil}, {Kim},
  {Klagyivik}, {Klar}, {Knude}, {Kochukhov}, {Kolka}, {Kos}, {Kutka}, {Lainey},
  {LeBouquin}, {Liu}, {Loreggia}, {Makarov}, {Marseille}, {Martayan},
  {Martinez-Rubi}, {Massart}, {Meynadier}, {Mignot}, {Munari}, {Nguyen},
  {Nordlander}, {Ocvirk}, {O'Flaherty}, {Olias Sanz}, {Ortiz}, {Osorio},
  {Oszkiewicz}, {Ouzounis}, {Palmer}, {Park}, {Pasquato}, {Peltzer}, {Peralta},
  {P{\'e}turaud}, {Pieniluoma}, {Pigozzi}, {Poels}, {Prat}, {Prod'homme},
  {Raison}, {Rebordao}, {Risquez}, {Rocca-Volmerange}, {Rosen}, {Ruiz-Fuertes},
  {Russo}, {Sembay}, {Serraller Vizcaino}, {Short}, {Siebert}, {Silva},
  {Sinachopoulos}, {Slezak}, {Soffel}, {Sosnowska}, {Strai{\v{z}}ys}, {ter
  Linden}, {Terrell}, {Theil}, {Tiede}, {Troisi}, {Tsalmantza}, {Tur},
  {Vaccari}, {Vachier}, {Valles}, {Van Hamme}, {Veltz}, {Virtanen}, {Wallut},
  {Wichmann}, {Wilkinson}, {Ziaeepour}, \& {Zschocke}}]{Gaia2}
{Gaia Collaboration}, {Prusti}, T., {de Bruijne}, J.~H.~J., {et~al.} 2016,
  \aap, 595, A1

\bibitem[{{Gallego-Cano} {et~al.}(2020){Gallego-Cano}, {Sch{\"o}del},
  {Nogueras-Lara}, {Dong}, {Shahzamanian}, {Fritz}, {Gallego-Calvente}, \&
  {Neumayer}}]{Laly_2020}
{Gallego-Cano}, E., {Sch{\"o}del}, R., {Nogueras-Lara}, F., {et~al.} 2020,
  \aap, 634, A71

\bibitem[{{Ginsburg} \& {Kruijssen}(2018)}]{cluster_formation_rate}
{Ginsburg}, A. \& {Kruijssen}, J.~M.~D. 2018, \apjl, 864, L17

\bibitem[{{Gordon} {et~al.}(2023){Gordon}, {de Witt}, \& {Jacobs}}]{SgA_pm}
{Gordon}, D., {de Witt}, A., \& {Jacobs}, C.~S. 2023, \aj, 165, 49

\bibitem[{{GRAVITY Collaboration} {et~al.}(2020){GRAVITY Collaboration},
  {Abuter}, {Amorim}, {Baub{\"o}ck}, {Berger}, {Bonnet}, {Brandner}, {Cardoso},
  {Cl{\'e}net}, {de Zeeuw}, {Dexter}, {Eckart}, {Eisenhauer}, {F{\"o}rster
  Schreiber}, {Garcia}, {Gao}, {Gendron}, {Genzel}, {Gillessen}, {Habibi},
  {Haubois}, {Henning}, {Hippler}, {Horrobin}, {Jim{\'e}nez-Rosales}, {Jochum},
  {Jocou}, {Kaufer}, {Kervella}, {Lacour}, {Lapeyr{\`e}re}, {Le Bouquin},
  {L{\'e}na}, {Nowak}, {Ott}, {Paumard}, {Perraut}, {Perrin}, {Pfuhl},
  {Rodr{\'\i}guez-Coira}, {Shangguan}, {Scheithauer}, {Stadler}, {Straub},
  {Straubmeier}, {Sturm}, {Tacconi}, {Vincent}, {von Fellenberg}, {Waisberg},
  {Widmann}, {Wieprecht}, {Wiezorrek}, {Woillez}, {Yazici}, \&
  {Zins}}]{GC_distance_gravity}
{GRAVITY Collaboration}, {Abuter}, R., {Amorim}, A., {et~al.} 2020, \aap, 636,
  L5

\bibitem[{{Habibi} {et~al.}(2014){Habibi}, {Stolte}, \&
  {Harfst}}]{Arches_Quint_tails}
{Habibi}, M., {Stolte}, A., \& {Harfst}, S. 2014, \aap, 566, A6

\bibitem[{{Harfst} {et~al.}(2010){Harfst}, {Portegies Zwart}, \&
  {Stolte}}]{Arches_mass}
{Harfst}, S., {Portegies Zwart}, S., \& {Stolte}, A. 2010, \mnras, 409, 628

\bibitem[{{Hosek} {et~al.}(2015){Hosek}, {Lu}, {Anderson}, {Ghez}, {Morris}, \&
  {Clarkson}}]{Arches_exctin}
{Hosek}, Matthew~W., J., {Lu}, J.~R., {Anderson}, J., {et~al.} 2015, \apj, 813,
  27

\bibitem[{{Hosek} {et~al.}(2019){Hosek}, {Lu}, {Anderson}, {Najarro}, {Ghez},
  {Morris}, {Clarkson}, \& {Albers}}]{Arches_IMF}
{Hosek}, Matthew~W., J., {Lu}, J.~R., {Anderson}, J., {et~al.} 2019, \apj, 870,
  44

\bibitem[{{Hosek} {et~al.}(2020){Hosek}, {Lu}, {Lam}, {Gautam}, {Lockhart},
  {Kim}, \& {Jia}}]{Spisea}
{Hosek}, Matthew~W., J., {Lu}, J.~R., {Lam}, C.~Y., {et~al.} 2020, \aj, 160,
  143

\bibitem[{Hosek {et~al.}(2022)Hosek, Do, Lu, Morris, Ghez, Martinez, \&
  Anderson}]{Hosek2022}
Hosek, M.~W., Do, T., Lu, J.~R., {et~al.} 2022, Measuring the Orbits of the
  Arches and Quintuplet Clusters using HST and Gaia: Exploring Scenarios for
  Star Formation Near the Galactic Center

\bibitem[{{Kroupa}(2001)}]{IMF}
{Kroupa}, P. 2001, \mnras, 322, 231

\bibitem[{{Kruijssen} {et~al.}(2014){Kruijssen}, {Longmore}, {Elmegreen},
  {Murray}, {Bally}, {Testi}, \& {Kennicutt}}]{cluster_dissolution}
{Kruijssen}, J.~M.~D., {Longmore}, S.~N., {Elmegreen}, B.~G., {et~al.} 2014,
  \mnras, 440, 3370

\bibitem[{Kunder {et~al.}(2012)Kunder, Koch, Rich, de~Propris, Howard, Stubbs,
  Johnson, Shen, Wang, Robin, Kormendy, Soto, Frinchaboy, Reitzel, Zhao, \&
  Origlia}]{Kunder_2012}
Kunder, A., Koch, A., Rich, R.~M., {et~al.} 2012, The Astronomical Journal,
  143, 57

\bibitem[{{Laporte} {et~al.}(2022){Laporte}, {Koposov}, \&
  {Belokurov}}]{filaments_mw}
{Laporte}, C. F.~P., {Koposov}, S.~E., \& {Belokurov}, V. 2022, \mnras, 510,
  L13

\bibitem[{{Launhardt} {et~al.}(2002{\natexlab{a}}){Launhardt}, {Zylka}, \&
  {Mezger}}]{NSD_size1}
{Launhardt}, R., {Zylka}, R., \& {Mezger}, P.~G. 2002{\natexlab{a}}, \aap, 384,
  112

\bibitem[{{Launhardt} {et~al.}(2002{\natexlab{b}}){Launhardt}, {Zylka}, \&
  {Mezger}}]{Lyman_flux}
{Launhardt}, R., {Zylka}, R., \& {Mezger}, P.~G. 2002{\natexlab{b}}, \aap, 384,
  112

\bibitem[{{Libralato} {et~al.}(2020){Libralato}, {Fardal}, {Lennon}, {van der
  Marel}, \& {Bellini}}]{LibralatoMC}
{Libralato}, M., {Fardal}, M., {Lennon}, D., {van der Marel}, R.~P., \&
  {Bellini}, A. 2020, \mnras, 497, 4733

\bibitem[{{Libralato} {et~al.}(2021){Libralato}, {Lennon}, {Bellini}, {van der
  Marel}, {Clark}, {Najarro}, {Patrick}, {Anderson}, {Bedin}, {Crowther}, {de
  Mink}, {Evans}, {Platais}, {Sabbi}, \& {Sohn}}]{LIBRALA2021}
{Libralato}, M., {Lennon}, D.~J., {Bellini}, A., {et~al.} 2021, \mnras, 500,
  3213

\bibitem[{{Liermann} {et~al.}(2012){Liermann}, {Hamann}, \&
  {Oskinova}}]{Quintuplet_4Myr}
{Liermann}, A., {Hamann}, W.~R., \& {Oskinova}, L.~M. 2012, \aap, 540, A14

\bibitem[{{Lu} {et~al.}(2013){Lu}, {Do}, {Ghez}, {Morris}, {Yelda}, \&
  {Matthews}}]{Lu_2013}
{Lu}, J.~R., {Do}, T., {Ghez}, A.~M., {et~al.} 2013, \apj, 764, 155

\bibitem[{{Mart{\'\i}nez-Arranz} {et~al.}(2022){Mart{\'\i}nez-Arranz},
  {Sch{\"o}del}, {Nogueras-Lara}, \& {Shahzamanian}}]{yo}
{Mart{\'\i}nez-Arranz}, {\'A}., {Sch{\"o}del}, R., {Nogueras-Lara}, F., \&
  {Shahzamanian}, B. 2022, \aap, 660, L3

\bibitem[{{Matsunaga} {et~al.}(2011){Matsunaga}, {Kawadu}, {Nishiyama},
  {Nagayama}, {Kobayashi}, {Tamura}, {Bono}, {Feast}, \&
  {Nagata}}]{three_cepheids}
{Matsunaga}, N., {Kawadu}, T., {Nishiyama}, S., {et~al.} 2011, \nat, 477, 188

\bibitem[{{Morris} \& {Serabyn}(1996)}]{Lyman_flux_III}
{Morris}, M. \& {Serabyn}, E. 1996, \araa, 34, 645

\bibitem[{{Najarro} {et~al.}(2004){Najarro}, {Figer}, {Hillier}, \&
  {Kudritzki}}]{Arches_2Myr}
{Najarro}, F., {Figer}, D.~F., {Hillier}, D.~J., \& {Kudritzki}, R.~P. 2004,
  \apjl, 611, L105

\bibitem[{{Nishiyama} {et~al.}(2008){Nishiyama}, {Nagata}, {Tamura}, {Kandori},
  {Hatano}, {Sato}, \& {Sugitani}}]{Lyman_flux_II}
{Nishiyama}, S., {Nagata}, T., {Tamura}, M., {et~al.} 2008, \apj, 680, 1174

\bibitem[{{Nishiyama} {et~al.}(2009){Nishiyama}, {Tamura}, {Hatano}, {Kato},
  {Tanab{\'e}}, {Sugitani}, \& {Nagata}}]{extinction_los}
{Nishiyama}, S., {Tamura}, M., {Hatano}, H., {et~al.} 2009, \apj, 696, 1407

\bibitem[{{Nogueras-Lara}(2022)}]{Noguera_solo}
{Nogueras-Lara}, F. 2022, \aap, 668, L8

\bibitem[{{Nogueras-Lara} {et~al.}(2018){Nogueras-Lara}, {Gallego-Calvente},
  {Dong}, {Gallego-Cano}, {Girard}, {Hilker}, {de Zeeuw}, {Feldmeier-Krause},
  {Nishiyama}, {Najarro}, {Neumayer}, \& {Sch{\"o}del}}]{GNSI}
{Nogueras-Lara}, F., {Gallego-Calvente}, A.~T., {Dong}, H., {et~al.} 2018,
  \aap, 610, A83

\bibitem[{{Nogueras-Lara} {et~al.}(2019{\natexlab{a}}){Nogueras-Lara},
  {Sch{\"o}del}, {Gallego-Calvente}, {Dong}, {Gallego-Cano}, {Shahzamanian},
  {Girard}, {Nishiyama}, {Najarro}, \& {Neumayer}}]{GNSII}
{Nogueras-Lara}, F., {Sch{\"o}del}, R., {Gallego-Calvente}, A.~T., {et~al.}
  2019{\natexlab{a}}, \aap, 631, A20

\bibitem[{{Nogueras-Lara} {et~al.}(2020){Nogueras-Lara}, {Sch{\"o}del},
  {Gallego-Calvente}, {Gallego-Cano}, {Shahzamanian}, {Dong}, {Neumayer},
  {Hilker}, {Najarro}, {Nishiyama}, {Feldmeier-Krause}, {Girard}, \&
  {Cassisi}}]{Paco_sfh}
{Nogueras-Lara}, F., {Sch{\"o}del}, R., {Gallego-Calvente}, A.~T., {et~al.}
  2020, Nature Astronomy, 4, 377

\bibitem[{{Nogueras-Lara} {et~al.}(2019{\natexlab{b}}){Nogueras-Lara},
  {Sch{\"o}del}, {Najarro}, {Gallego-Calvente}, {Gallego-Cano}, {Shahzamanian},
  \& {Neumayer}}]{paco_exctinction}
{Nogueras-Lara}, F., {Sch{\"o}del}, R., {Najarro}, F., {et~al.}
  2019{\natexlab{b}}, \aap, 630, L3

\bibitem[{{Nogueras-Lara} {et~al.}(2021){Nogueras-Lara}, {Sch{\"o}del}, \&
  {Neumayer}}]{GNSIV}
{Nogueras-Lara}, F., {Sch{\"o}del}, R., \& {Neumayer}, N. 2021, \aap, 653, A133

\bibitem[{{Nogueras-Lara} {et~al.}(2022){Nogueras-Lara}, {Sch{\"o}del}, \&
  {Neumayer}}]{Paco_SgrB2}
{Nogueras-Lara}, F., {Sch{\"o}del}, R., \& {Neumayer}, N. 2022, Nature
  Astronomy, 6, 1178

\bibitem[{Portegies~Zwart {et~al.}(2001)Portegies~Zwart, Makino, McMillan, \&
  Hut}]{dissolve_GC}
Portegies~Zwart, S., Makino, J., McMillan, S., \& Hut, P. 2001, The
  Astrophysical Journal, 565

\bibitem[{{Rui} {et~al.}(2019){Rui}, {Hosek}, {Lu}, {Clarkson}, {Anderson},
  {Morris}, \& {Ghez}}]{Quintuplet_hl}
{Rui}, N.~Z., {Hosek}, Matthew~W., J., {Lu}, J.~R., {et~al.} 2019, \apj, 877,
  37

\bibitem[{{Sander} {et~al.}(1998){Sander}, {Ester}, {Kriegel}, \&
  {Xu}}]{dbscan_original}
{Sander}, J., {Ester}, M., {Kriegel}, H.-P., \& {Xu}, X. 1998, Data Mining and
  Knowledge Discovery, 2, 169

\bibitem[{{Schoedel} {et~al.}(2014){Schoedel}, {Feldmeier}, {Kunneriath},
  {Stolovy}, \& {Neumauer}}]{rainer_2014}
{Schoedel}, R., {Feldmeier}, A., {Kunneriath}, D., {Stolovy}, S., \&
  {Neumauer}, N. 2014, VizieR Online Data Catalog, J/A+A/566/A47

\bibitem[{Schonrich {et~al.}(2015)Schonrich, Aumer, \& Sale}]{Sch2015}
Schonrich, R., Aumer, M., \& Sale, S.~E. 2015, The Astrophysical Journal, 812,
  L21

\bibitem[{Schubert {et~al.}(2017)Schubert, Sander, Ester, Kriegel, \&
  Xu}]{dbscan2}
Schubert, E., Sander, J., Ester, M., Kriegel, H.~P., \& Xu, X. 2017, ACM Trans.
  Database Syst., 42

\bibitem[{{Shahzamanian} {et~al.}(2019){Shahzamanian}, {Sch{\"o}del},
  {Nogueras-Lara}, {Dong}, {Gallego-Cano}, {Gallego-Calvente}, \&
  {Gardini}}]{Ban_cluster}
{Shahzamanian}, B., {Sch{\"o}del}, R., {Nogueras-Lara}, F., {et~al.} 2019,
  \aap, 632, A116

\bibitem[{{Shahzamanian} {et~al.}(2022){Shahzamanian}, {Sch{\"o}del},
  {Nogueras-Lara}, {Mart{\'\i}nez-Arranz}, {Sormani}, {Gallego-Calvente},
  {Gallego-Cano}, \& {Alburai}}]{Ban_catalog}
{Shahzamanian}, B., {Sch{\"o}del}, R., {Nogueras-Lara}, F., {et~al.} 2022,
  \aap, 662, A11

\bibitem[{{Sormani} {et~al.}(2022){Sormani}, {Sanders}, {Fritz}, {Smith},
  {Gerhard}, {Sch{\"o}del}, {Magorrian}, {Neumayer}, {Nogueras-Lara},
  {Feldmeier-Krause}, {Mastrobuono-Battisti}, {Schultheis}, {Shahzamanian},
  {Vasiliev}, {Klessen}, {Lucas}, \& {Minniti}}]{Sormani2022}
{Sormani}, M.~C., {Sanders}, J.~L., {Fritz}, T.~K., {et~al.} 2022, \mnras, 512,
  1857

\bibitem[{{Speagle}(2020)}]{Dynesty}
{Speagle}, J.~S. 2020, \mnras, 493, 3132

\bibitem[{{Stolker} {et~al.}(2020){Stolker}, {Quanz}, {Todorov}, {K{\"u}hn},
  {Molli{\`e}re}, {Meyer}, {Currie}, {Daemgen}, \& {Lavie}}]{species}
{Stolker}, T., {Quanz}, S.~P., {Todorov}, K.~O., {et~al.} 2020, \aap, 635, A182

\bibitem[{{Stolovy} {et~al.}(2006){Stolovy}, {Ramirez}, {Arendt}, {Cotera},
  {Yusef-Zadeh}, {Law}, {Gezari}, {Sellgren}, {Karr}, {Moseley}, \&
  {Smith}}]{picture_2}
{Stolovy}, S., {Ramirez}, S., {Arendt}, R.~G., {et~al.} 2006, in Journal of
  Physics Conference Series, Vol.~54, Journal of Physics Conference Series,
  176--182

\bibitem[{{Stolte} {et~al.}(2008){Stolte}, {Ghez}, {Morris}, {Lu}, {Brandner},
  \& {Matthews}}]{Arches_dispersion}
{Stolte}, A., {Ghez}, A.~M., {Morris}, M., {et~al.} 2008, \apj, 675, 1278

\bibitem[{{Stolte} {et~al.}(2014){Stolte}, {Hu{\ss}mann}, {Morris}, {Ghez},
  {Brandner}, {Lu}, {Clarkson}, {Habibi}, \&
  {Matthews}}]{Quintuplet_dispersion}
{Stolte}, A., {Hu{\ss}mann}, B., {Morris}, M.~R., {et~al.} 2014, \apj, 789, 115

\bibitem[{{Virtanen} {et~al.}(2020){Virtanen}, {Gommers}, {Oliphant},
  {Haberland}, {Reddy}, {Cournapeau}, {Burovski}, {Peterson}, {Weckesser},
  {Bright}, {van der Walt}, {Brett}, {Wilson}, {Millman}, {Mayorov}, {Nelson},
  {Jones}, {Kern}, {Larson}, {Carey}, {Polat}, {Feng}, {Moore}, {VanderPlas},
  {Laxalde}, {Perktold}, {Cimrman}, {Henriksen}, {Quintero}, {Harris},
  {Archibald}, {Ribeiro}, {Pedregosa}, {van Mulbregt}, \& {SciPy 1. 0
  Contributors}}]{Scipy}
{Virtanen}, P., {Gommers}, R., {Oliphant}, T.~E., {et~al.} 2020, Nature
  Methods, 17, 261

\end{thebibliography}
	
	%
		
\begin{onecolumn}
\begin{appendix}

	\section{Quality Check}
		\label{quality_check}
	To assess the quality of the data, we identified the NSD  and Bulge through stellar kinematics and compared the obtained values with those reported in the literature. Firstly, we transformed the proper motions from equatorial to Galactic with the package \textit{SkyCoord} from \textit{astropy} \citep{astropy}. Since the proper motions in the L21 catalog are in the Gaia DR2 reference frame, we further transformed them into a reference frame where SgrA* is at rest. This transformation involved subtracting the velocity of SgrA* in the International Celestial Reference Frame (ICRF), which is ($\mu_{l}, \mu_{b}$)$^{SgrA*}$ = $-$6.40, $-$0.24 mas/yr \citep{SgA_pm}. In Fig. \ref{nsd_fit} we can see the distribution of the Galactic proper motions of L21 for the components perpendicular and parallel to the Galactic plane (gray histograms). Then, we fit different Gaussian models to these distributions using the python package \textit{dynesty} \citep{Dynesty}. We found that a two-Gaussians fit best reproduces the perpendicular component and three the parallel one (see Fig. \ref{nsd_fit} and Fig.\ref{corner_plots}). In Tab. \ref{tablefit} we can see the values for these Gaussians, which we interpret as representative of the Bulge and NSD populations
	(see \citealt{Ban_catalog}). In the left panel of Fig. \ref{nsd_fit}, the red Gaussian represents the Bulge population and the black one the NSD. In the right panel, the red Gaussian also represents the Bulge population. The blue one represents the stars of the NSD that stream towards the Galactic East and the black one those that stream towards the Galactic West. The Bulge velocity in this reference frame should ideally be zero, but we can see that the parallel component is $\mu_{l}$ = 0.64 mas/yr. Due to data incompleteness, we tend to detect more stars from the near side of the NSD, introducing a bias in velocities towards stars moving to the West. To rectify this bias, we adjusted the Bulge component to center it around zero. Consequently, the revised values for the NSD components are $\mu_{l}$ = 1.97 mas/yr and $\mu_{l}$ = -2.17 mas/yr. These revised results are consistent, within the known uncertainties, with the values previously determined for the mean velocities of stars in the NSD \citep{Kunder_2012,Sch2015,Ban_catalog,Sormani2022,yo,Noguera_solo}. It is noteworthy to mention that in \cite{LIBRALA2021}, the fitting of the data for the parallel component solely involves the use of two Gaussians, without considering the existence of the NSD.

	\begin{figure}[!h]
	\centering

	\subfloat[]{
		\includegraphics[width=0.45\textwidth]{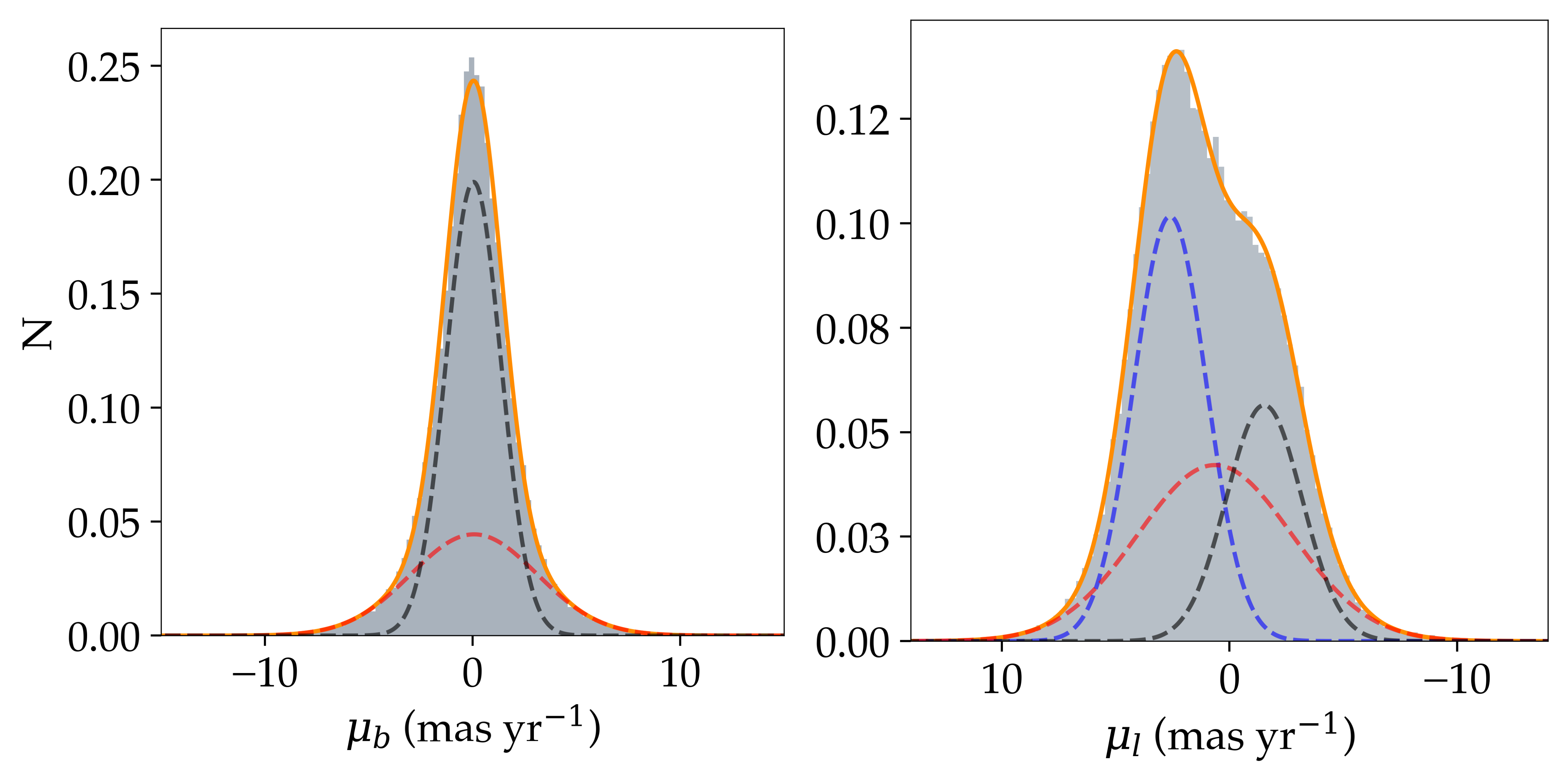}
		\label{nsd_fit}
	}
	\quad
	\subfloat[]{
		\begin{tabular}{@{}lllc@{}}
			\toprule
			\multicolumn{1}{c}{Perpendicular}  & \multicolumn{1}{c}{Bulge} & \multicolumn{1}{c}{NSD} &                                      \\ \midrule
			$\mu_{b}\,(mas\,yr^{-1})$          & 0.07 $\pm$ 0.04          & 0.05 $\pm$ 0.02        & -                                    \\
			$\sigma_{\mu_{b}}\,(mas\,yr^{-1})$ & 3.04 $\pm$ 0.05           & 1.32 $\pm$ 0.02         & -                                    \\
			$amp_{b}$                          & 0.34 $\pm$ 0.02           & 0.65 $\pm$ 0.02         & -                                    \\ \midrule
			\multicolumn{1}{c}{Parallel}       & \multicolumn{1}{c}{Bulge} & \multicolumn{1}{c}{NSD} & NSD                                  \\ \midrule
			$\mu_{l}\,(mas\,yr^{-1})$          & 0.64$\pm$0.07          & 2.61 $\pm$ 0.07      & \multicolumn{1}{l}{-1.53 $\pm$ 0.13} \\
			$\sigma_{\mu_{l}}\,(mas\,yr^{-1})$ & 3.38$\pm$0.13             & 1.60 $\pm$ 0.07         & \multicolumn{1}{l}{1.66 $\pm$ 0.09}  \\
			$amp_{l}$                          & 0.36 $\pm$ 0.07           & 0.41 $\pm$ 0.04         & \multicolumn{1}{l}{0.23 $\pm$ 0.04}  \\ \bottomrule
			\label{tablefit}
		\end{tabular}
	}
	\caption{(a): Gray histograms represent the proper motion distributions for the perpendicular (left) and parallel (right) proper motion components of the stars in L21. The red Gaussian represents the bulge stars in both plots. The black Gaussian on the left represents the perpendicular component of the NSD stars. On the right, the blue and black Gaussians represent stars in the near and far side of the NSD respectively. (b): Best-fit parameters and uncertainties for L21 data (Fig. \ref{nsd_fit}). $\mu$, $\sigma$, and $amp$ represent the mean velocity, standard deviation, and amplitude of the Gaussians fitted to the distribution.}
\end{figure}

\begin{figure}[h!]
	\begin{subfigure}{0.50\linewidth}
		\includegraphics[width=\linewidth]{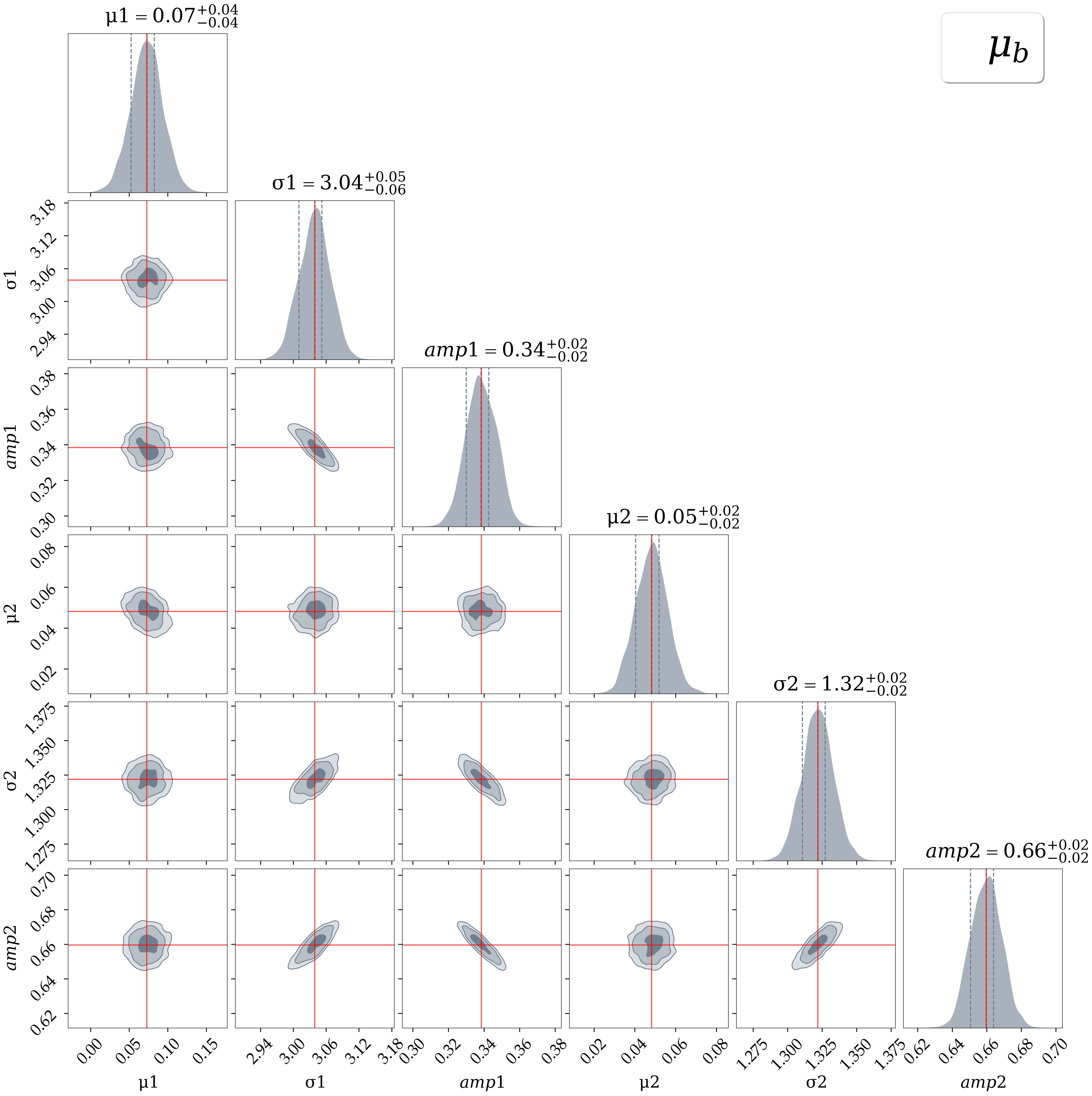}
	\end{subfigure}
	\hfill 
	\begin{subfigure}{0.50\linewidth}
		\includegraphics[width=\linewidth]{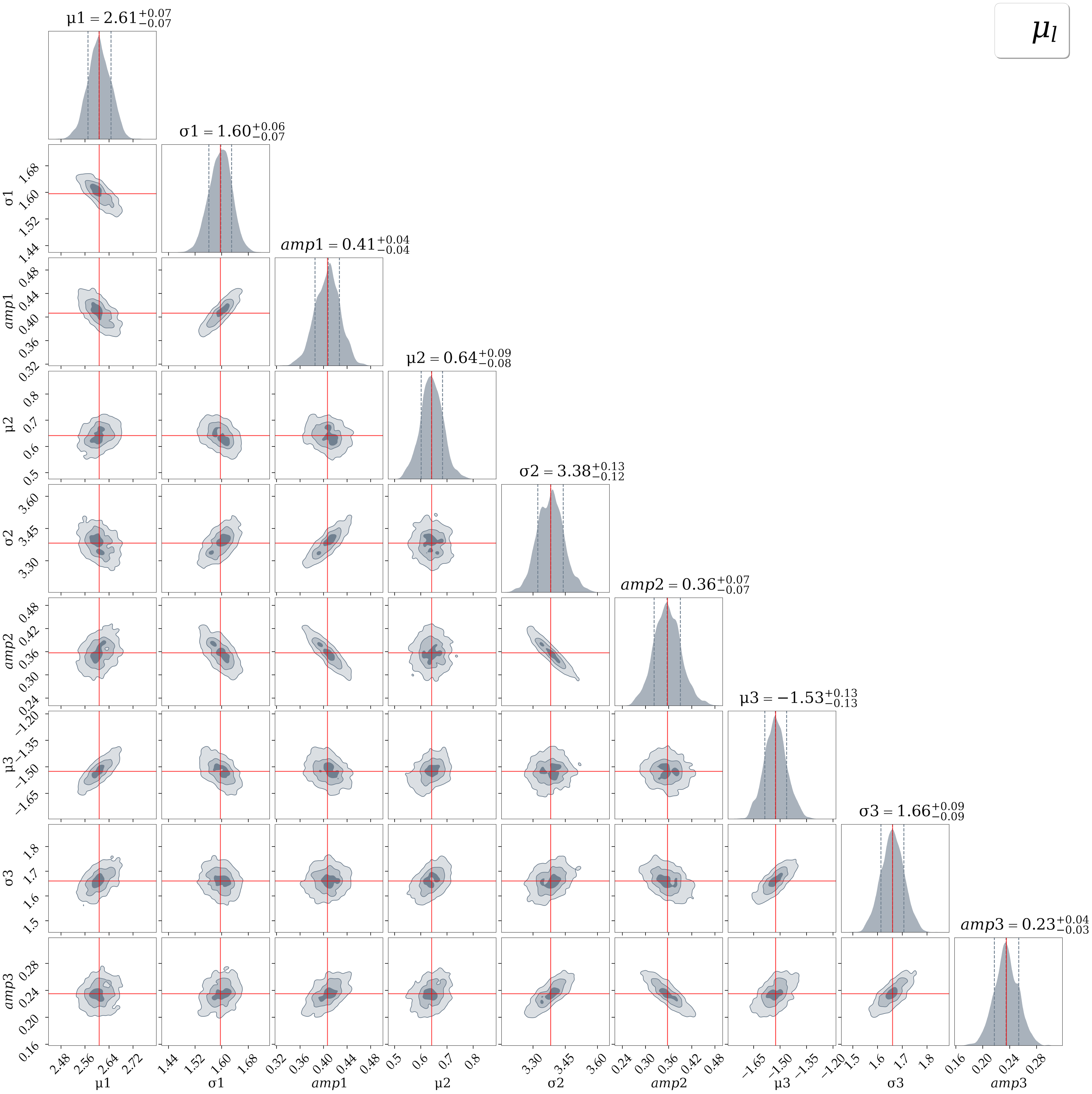}
	\end{subfigure}
	\caption{Posterior probability distributions of the free parameters for  Gaussian fitting. Left for the perpendicular component of the proper motions and right for the parallel one. }
	\label{corner_plots}
\end{figure}
\FloatBarrier
\section{Testing the algorithm}
\begin{figure*}[!h]	
	\begin{subfigure}{0.49\textwidth}
		\includegraphics[width=\textwidth]{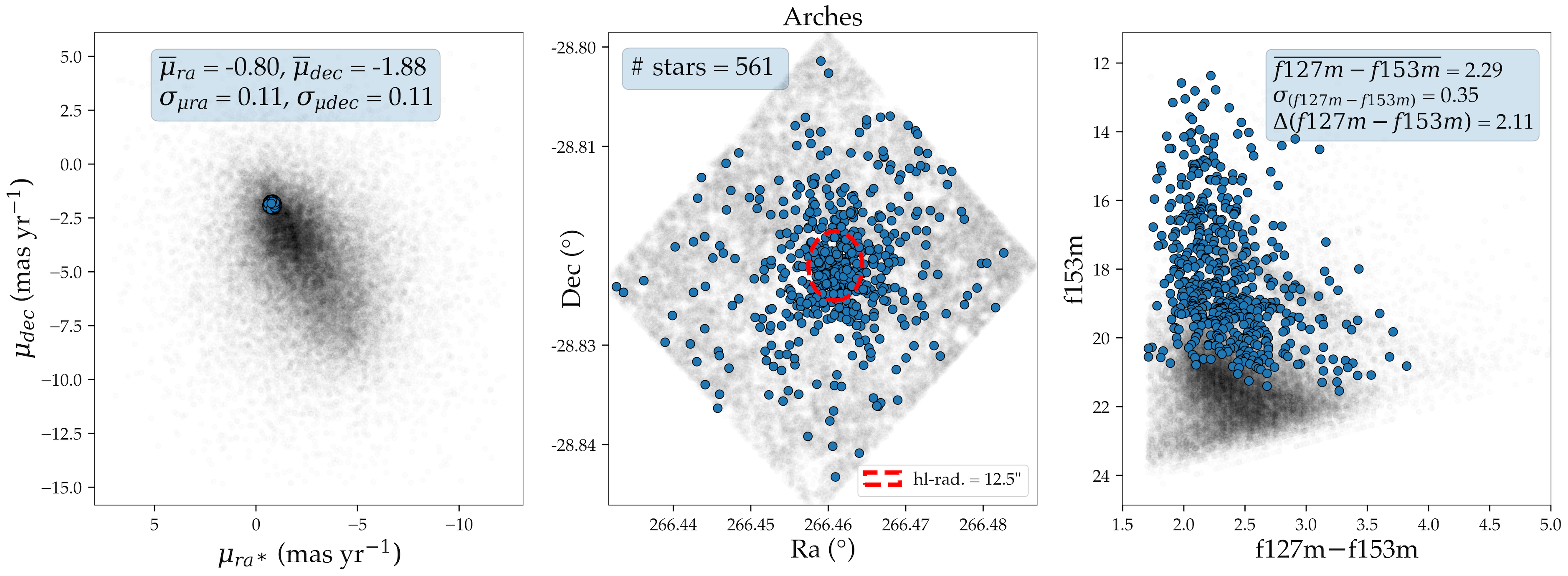}
	\end{subfigure}
	\hfill
	\begin{subfigure}[!h]{0.49\textwidth}
		\includegraphics[width=\textwidth]{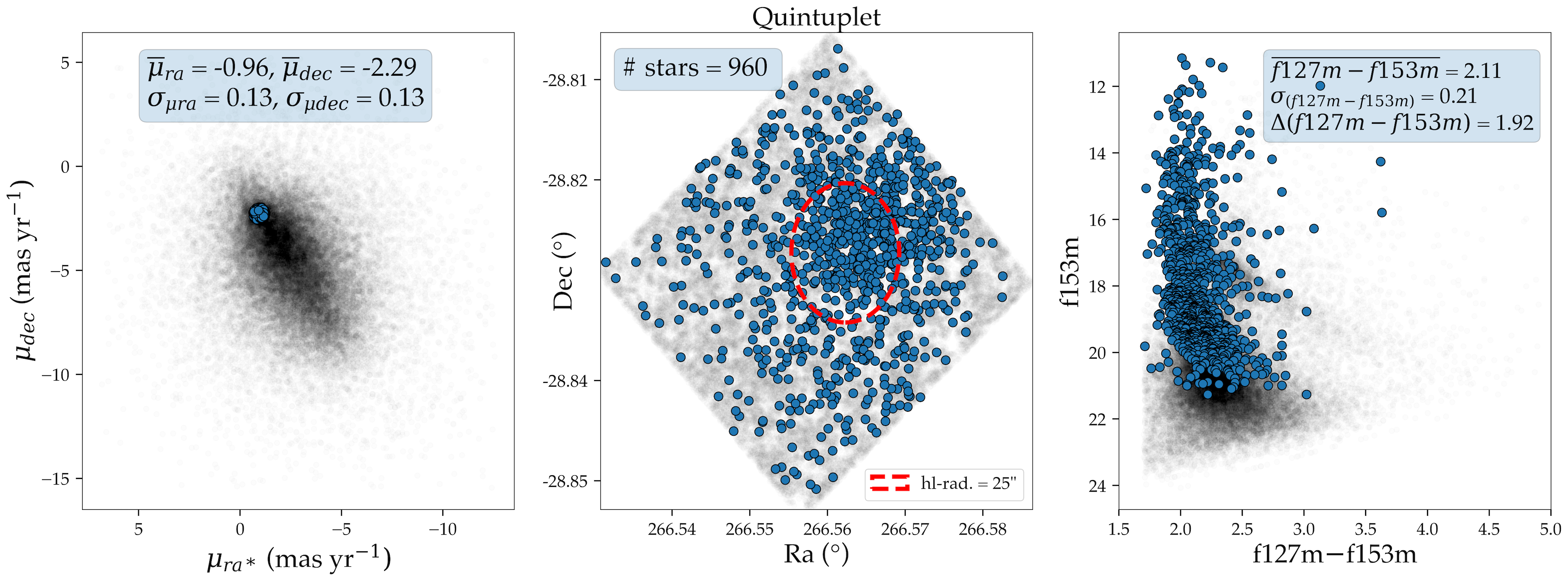}
	\end{subfigure}

	\begin{subfigure}{0.49\textwidth}
	\includegraphics[width=\textwidth]{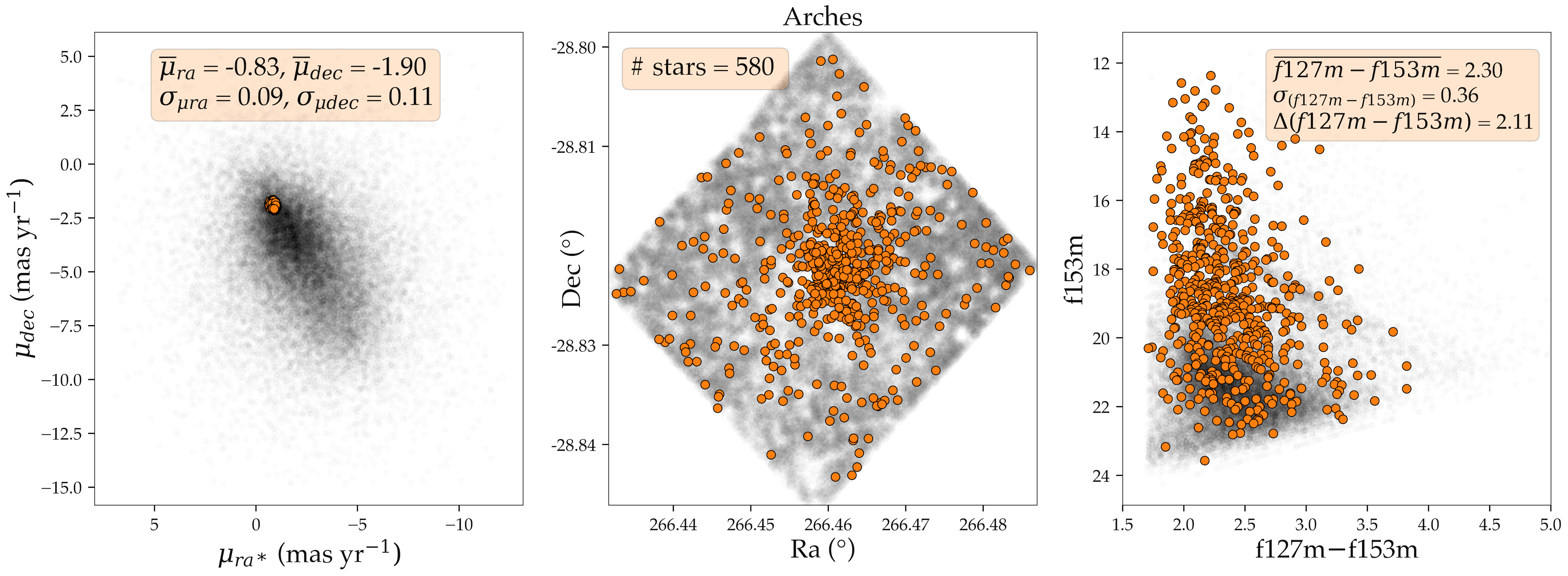}
\end{subfigure}
\hfill
\begin{subfigure}[!h]{0.49\textwidth}
	\includegraphics[width=\textwidth]{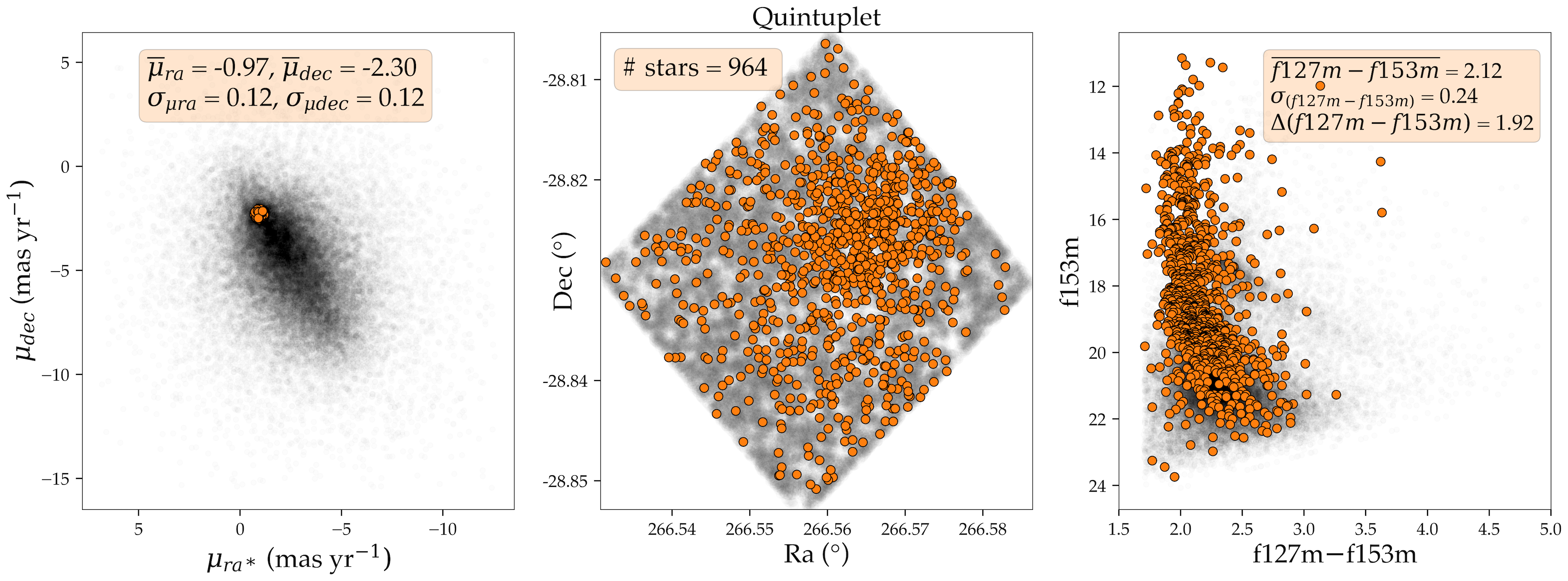}
\end{subfigure}
\caption{In the top row, we present the members of the Arches and Quintuplet clusters according to \cite{Hosek2022}. The blue points indicate stars with a likelihood $\geq$ 0.7 of belonging to the respective clusters, as determined in the aforementioned study. The red circles represent the half-light radii for each cluster, as found in the literature \citep{Arches_exctin, Quintuplet_hl}. The left columns display the vector-point diagram, the middle columns show the stellar positions, and the right columns display the CMD. In the bottom row, we present the Arches and Quintuplet clusters as recovered by the algorithm when configured to search only in 2D space formed by  the proper motion components. With this configuration, 75\% of the stars labeled as Arches members match those of \cite{Hosek2022}, and 85\% in the case of Quintuplet cluster. }
	\label{hos_prob}
\end{figure*}
\FloatBarrier
  \newpage
	\section{Results}
		\newcommand{\scale}{1}

	\begin{figure}[!h]
	\centering
		\begin{subfigure}{\scale\textwidth}
			\includegraphics[width=\linewidth]{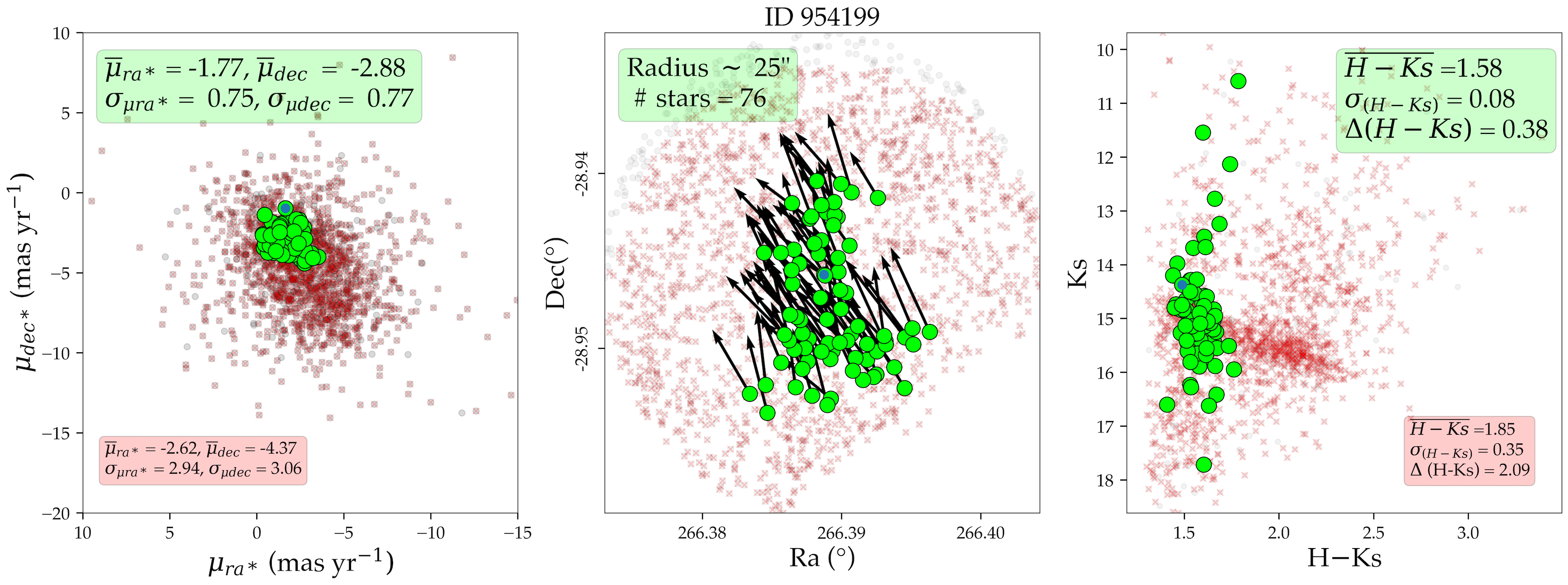}
		
			\label{ms_2}
		\end{subfigure}
	
%

		\begin{subfigure}{\scale\textwidth}
			\includegraphics[width=\linewidth]{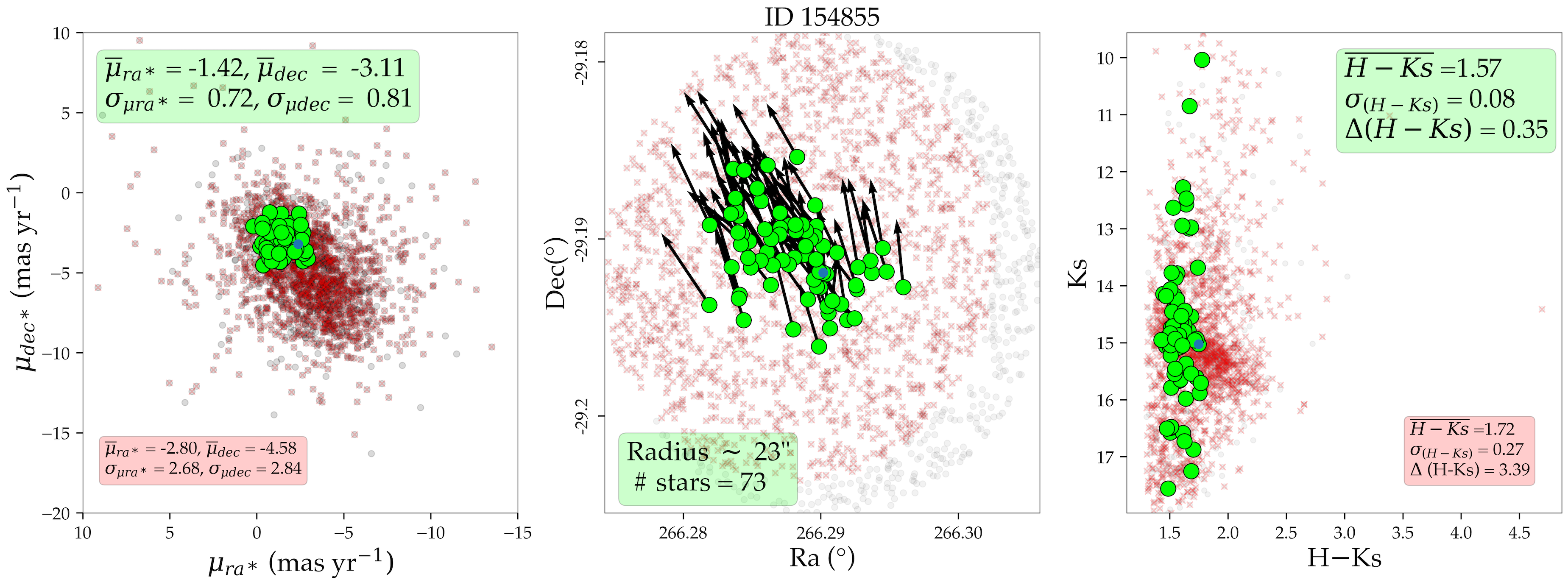}
			
			\label{ms_4}
		\end{subfigure}
 
		\begin{subfigure}{\scale\textwidth}
			\includegraphics[width=\linewidth]{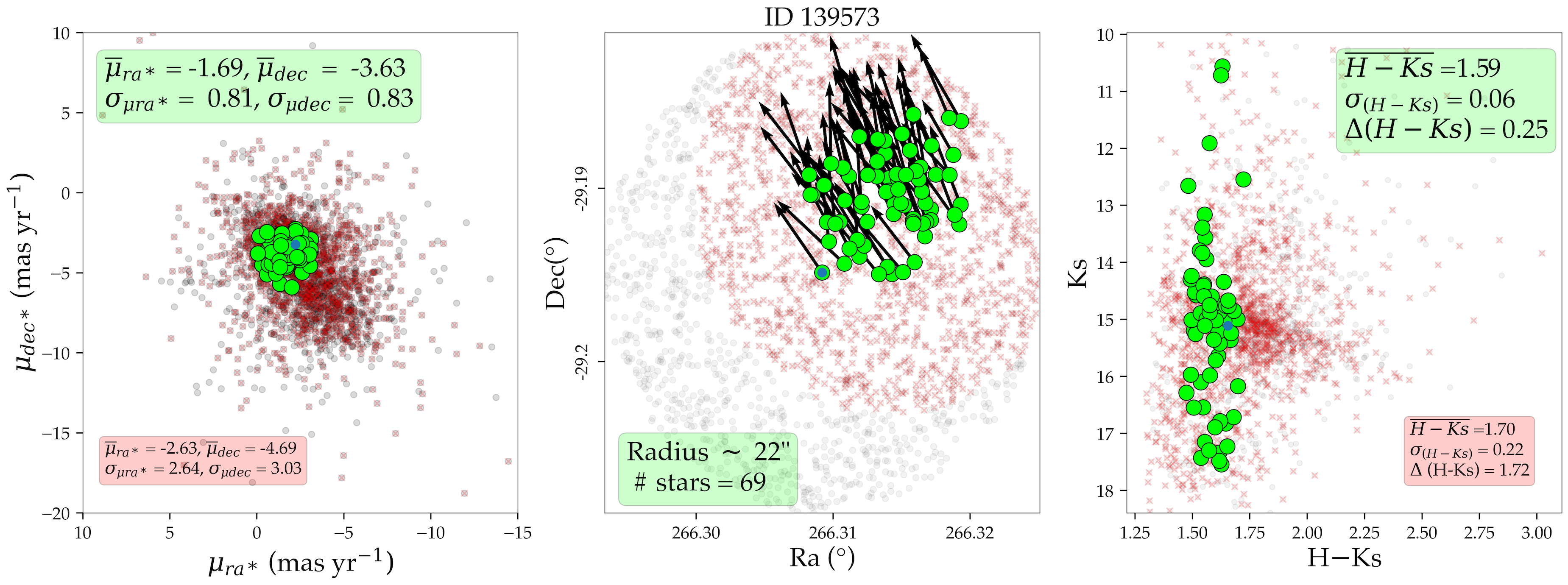}
		
			\label{ms_5}
		\end{subfigure}

		\caption{Same plots as in Fig. \ref{ms_1} for the rest of massive stars  that belong to  a co-moving group (green points in Fig. \ref{region}) }
		\label{ms_all}
	\end{figure}


%

\end{appendix}
\end{onecolumn}
	\end{document}